\documentclass[12pt,preprint, showpacs,preprintnumbers,aps,pre,groupedaddress]{revtex4-1}

\usepackage{graphicx}
\usepackage{subfigure}
\usepackage{color}
\usepackage{amsmath}
\usepackage[linktocpage,colorlinks=true,linkcolor=blue,citecolor=blue,breaklinks=true]{hyperref}
\usepackage{verbatim}
\usepackage{breakcites}
\usepackage{enumerate}

\begin{document}


\title{Driven active and passive nematics}

 
 \author{Sumesh P. Thampi}
\affiliation{The Rudolf Peierls Centre for Theoretical Physics, 1 Keble Road, Oxford, OX1 3NP, UK}

\author{Ramin Golestanian}
\affiliation{The Rudolf Peierls Centre for Theoretical Physics, 1 Keble Road, Oxford, OX1 3NP, UK}

\author{Julia M. Yeomans}
\affiliation{The Rudolf Peierls Centre for Theoretical Physics, 1 Keble Road, Oxford, OX1 3NP, UK}
\email[]{j.yeomans1@physics.ox.ac.uk}
\homepage[]{http://www-thphys.physics.ox.ac.uk/people/JuliaYeomans/}

\begin{abstract}
We investigate similarities in the micro-structural dynamics between externally driven and actively driven nematics. Walls, lines of strong deformations in the director field, and topological defects are characteristic features of an active nematic. Similar structures form in driven passive nematics when there are inhomogeneities in imposed velocity gradients due to non-linear flow fields or geometrical constraints. Specifically, pressure driven flow of a tumbling passive nematic in an expanding-contracting channel produces walls and defects similar to those seen in active nematics. We also study the response of active nematics to external driving, confirming that imposed shear suppresses the hydrodynamic instabilities. We show that shear fields can lead to  wall alignments and the localisation of active turbulence.
\bigskip

\end{abstract}

\maketitle

%
%
%
\section{Introduction}

Active matter is composed of entities that produce their own energy and hence active systems are continually driven out of thermodynamic equilibrium \cite{Marchetti2013, Ganesh2011, Sriram2010}. Biological examples, across a range of length scales, include actin or microtubule filaments powered by molecular motors \cite{Bausch2013b, Dogic2012}, bacterial suspensions \cite{Julia2012, Aranson2012} and flocks of birds \cite{Cavagna2014}. Moreover self-propelled particles can be fabricated to move by exploiting phoretic forces \cite{Masoud2014,Ramin2012}. Vertically vibrated granular rods are also often categorised as active systems \cite{Narayan2007}.

\begin{figure}
\center
 \subfigure[]{\includegraphics[trim = 90 30 70 20, clip, width=0.4\linewidth]{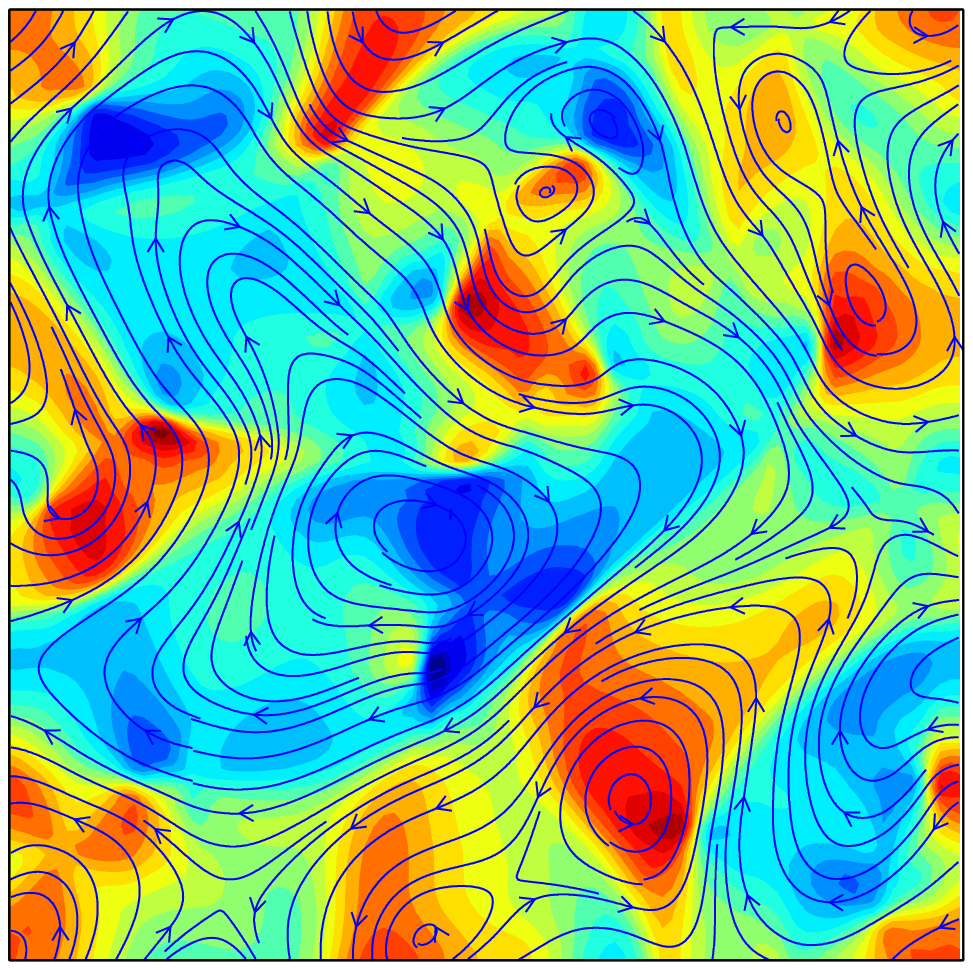}}
  \subfigure[]{\includegraphics[trim = 90 35 70 20, clip, width=0.44\linewidth]{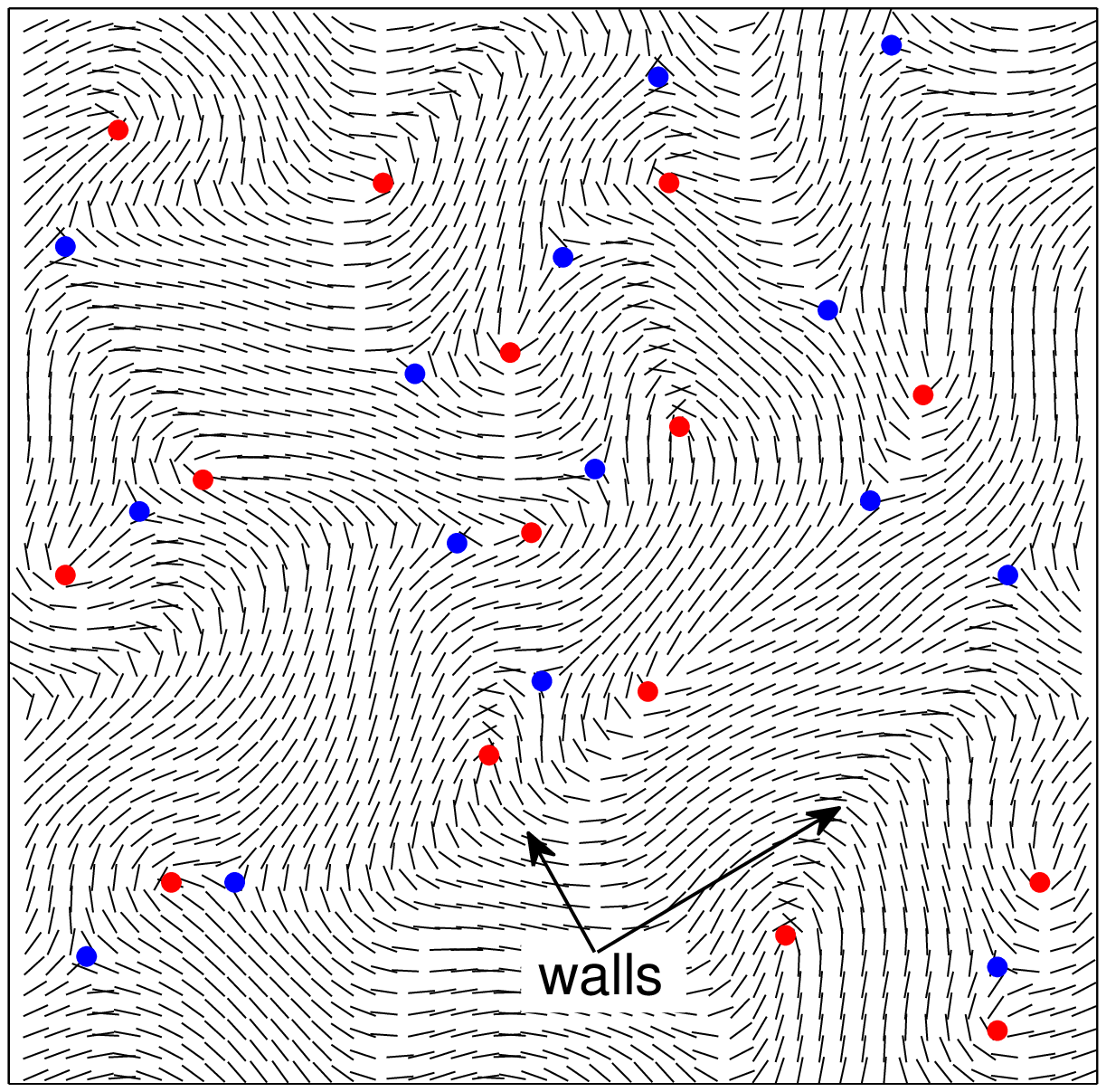}}
  \caption{Snapshots of the active turbulent state. (a) Flow field. The continuous lines are streamlines and the colour shading represents the vorticity field with blue and red shades respectively indicating clockwise and anticlockwise vorticity. (b) Corresponding nematic director field. Topological defects of charge $\pm {1}/{2}$ are  shown by red and blue markings respectively. Lines of strong deformations, which we shall term walls, are also distinctly visible in the director field.}
  \label{fig:actturb}
 \end{figure}

Many active systems have nematic symmetry and can be described by the same hydrodynamic equations of motion as passive nematic liquid crystals, but with an additional active term in the stress tensor. The active stress generates a flow field whenever there is a gradient in the magnitude or direction of the nematic order. As a result, active nematics are hydrodynamically unstable, and dense active suspensions exhibit chaotic turbulent-like flow states known as active turbulence \cite{Julia2012, Aranson2012, Giomi2014arxiv}. Active turbulence is characterised by regions of strong vorticity in the velocity field as shown in Fig.~\ref{fig:actturb}(a). 

Fig.~\ref{fig:actturb}(b) shows a typical corresponding director field. The hydrodynamic instabilities generate \textit{walls}, lines of strong distortions in the director field \cite{ourepl2014, Giomi2014}. In an extensile active nematic, these walls are bend distortions as indicated in Fig.~\ref{fig:actturb}(b). These walls separate nematically ordered domains, and  
are sources of hydrodynamic stress that contribute to the chaotic flow fields shown in Fig.~\ref{fig:actturb}. However the walls are not stable structures. They disintegrate into one or more pairs of oppositely charged ($\pm \frac{1}{2}$) topological defects. Elastic forces and flow drive the defects apart, and they subsequently move around until oppositely charged defects meet and annihilate. Both defect creation and annihilation events regenerate nematic regions which again undergo instabilities leading to wall formation and providing a mechanism that sustains active turbulence. However, despite this qualitative understanding, a comprehensive theory describing the dynamics of walls and defects is yet to emerge. 

Active liquid crystals operate out of equilibrium because of their intrinsic activity which leads to an input of energy at every point in the fluid. However, nematics, both passive and active, can also be driven out of equilibrium by an applied shear or pressure gradient. In this paper we numerically solve the equations of motion of both active and passive nematics under shear and pressure-driven flow and compare the effects of external and internal driving. In particular we show that, in certain circumstances, the walls and topological defects characteristic of active turbulence can also be a feature of driven, passive nematics.

Firstly, in Secs.~\ref{sec:theory} and \ref{sec:numerics}, we summarise the relevant hydrodynamic equations and the method that we use to solve them. In Sec.~\ref{sec:passive} we discuss the effect of shear, confinement and Poiseuille flow on 2D passive nematics, concentrating in particular on the tumbling regime. We show that pressure driven flow of a tumbling nematic suspension leads to walls, which generate topological defects in a channel of non-uniform cross section. In Sec.~\ref{sec:active} we turn to active nematics.  Linear stability analysis has shown that there is a threshold shear beyond which active hydrodynamic instabilities are suppressed for aligning suspensions \cite{Muhuri2007}. Solving the nonlinear equations we verify this analysis and show that this suppression of activity is observed in the tumbling regime as well. A corollary is that
in Poiseuille flow, active turbulence is increasingly confined to the centre of the channel as the flow velocity increases. Sec.~\ref{sec:discussion} concludes the paper with a discussion of the relevance of our analysis of passively driven systems to active systems.

%
\section{Equations of Motion}
\label{sec:theory}

The equations of nemato-hydrodynamics are described in this section. These are applicable to both passive and active nematics, except for having an additional stress term when used to describe the latter. Therefore the system under consideration may be passive, a molecular or colloidal liquid crystal, or active, such as a suspension of bacteria or microtubule bundles driven by molecular motors \cite{Marchetti2013}.

Both density and momentum are conserved variables, while orientational order is not. As well known for nematic liquid crystals, we use a second order traceless symmetric tensor  $\mathbf{Q}$ to describe the orientational order. It is defined as $\mathbf{Q} = \frac{q}{2} ( 3\mathbf{nn} - \mathbf{I})$ where $q$ is the largest eigenvalue of $\mathbf{Q}$ and is a measure of nematic degree of order, $\mathbf{n}$ is the director field and $\mathbf{I}$ is the identity tensor. We shall solve the equations of motion describing the evolution of momentum  $\rho \mathbf{u}$ and order parameter $\mathbf{Q}$ \cite{DeGennesBook, Berisbook} 
\begin{align}
\rho (\partial_t + u_k \partial_k) u_i = \partial_j \Pi_{ij},
\label{eqn:ns} \\
(\partial_t + u_k \partial_k) Q_{ij} - S_{ij} = D H_{ij},
\label{eqn:lc}
\end{align}
along with the equation of continuity. 

In addition to the advection, the response of $\mathbf{Q}$ to  velocity gradients of the flow field is also taken into account by the generalised advection term 
\begin{align}
S_{ij} = (\lambda E_{ik} + \Omega_{ik})(Q_{kj} + \delta_{kj}/3) + (Q_{ik} + \delta_{ik}/3)&(\lambda E_{kj} - \Omega_{kj})\nonumber\\
 &- 2 \lambda (Q_{ij} + \delta_{ij}/3)(Q_{kl}\partial_k u_l),
\end{align}
where $E_{ij} = (\partial_i u_j + \partial_j u_i)/2$ is the strain rate tensor and $\Omega_{ij} = (\partial_j u_i - \partial_i u_j)/2$ is the vorticity tensor. 

Vorticity $\boldsymbol{\omega} = \nabla \times \mathbf{u}$, the curl of velocity field describing the local rotation of the fluid elements, is related to the vorticity tensor by $\epsilon_{ijk}\omega_k = -2\Omega_{ij}$ where $\boldsymbol{\epsilon}$ is the Levi-Civita symbol. We will find that gradients of vorticity plays an important role in generating nematic director textures.

The competition between elongation and rotation is determined by the alignment parameter $\lambda$ which describes the aligning/tumbling motion of the director field in response to a simple shear flow in both passive and active nematics. From a microscopic perspective $\lambda$ is determined by the degree of orientational order and the shape of the constituent particles. The nematic suspension is usefully characterised in terms of $\lambda_1 = (3q+4)\lambda/9q$. For a tumbling material, $\lambda_1 < 1$, the director undergoes continuous rotation in response to a simple shear flow. Instead, for $\lambda_1 > 1$ the director aligns at a fixed angle, known as the Leslie angle $\theta_l = \frac{1}{2}\cos^{-1}({1}/{\lambda_1})$, with respect to flow direction \cite{Rienacker2002, Berisbook, DeGennesBook}.

We use the Landau-de Gennes free energy containing  a gradient energy term and bulk energy terms, 
\begin{align}
\mathcal{F} = K (\partial_k Q_{ij})^2/2 + A Q_{ij} Q_{ji}/2 + B Q_{ij} Q_{jk} Q_{ki}/3 + C (Q_{ij} Q_{ji})^2/4,
\end{align}
to prescribe the equilibrium of the order parameter $\mathbf{Q}$. Here, $A, B$ and $C$ are material constants that determine the conditions for the isotropic-nematic phase transition. We use the single elastic constant approximation where the elastic constant $K$ assigns the same energy cost to splay, bend and twist deformations. The variational derivative of $\mathcal{F}$ with respect to $\mathbf{Q}$ defines the molecular field $H_{ij} = -\delta \mathcal{F}/ \delta Q_{ij} + (\delta_{ij}/3) {\rm Tr} (\delta \mathcal{F}/ \delta Q_{kl})$. The symmetric, traceless matrix $\mathbf{H}$ describes the relaxation of $\mathbf{Q}$ in Eq.~(\ref{eqn:lc}) at a rate controlled by the rotational diffusivity, $D$.

The stresses acting on a fluid element may be classified into
\begin{enumerate}[i)]
\item the viscous stress, $\Pi_{ij}^{viscous} = 2 \mu E_{ij}$,  
\item the elastic stress,
\begin{align}
\Pi_{ij}^{passive}=-P\delta_{ij} + 2 \lambda(Q_{ij} + \delta_{ij}/3) (Q_{kl} H_{lk})
-\lambda H_{ik} (Q_{kj} + \delta_{kj}/3) \nonumber\\ - \lambda (Q_{ik} + \delta_{ik}/3) H_{kj}
-\partial_i Q_{kl} \frac{\delta \mathcal{F}}{\delta \partial_j Q_{lk}} + Q_{ik}H_{kj} - H_{ik} Q_{kj},
\end{align}
%
\item the active stress, $\Pi_{ij}^{active} = -\zeta Q_{ij}$,
\end{enumerate}
where the modified bulk pressure is $P = \rho T - \frac{K}{2} (\partial_k Q_{ij})^2$ and the Newtonian viscosity of the suspension is $\mu$. The elastic stress term, $\Pi_{ij}^{passive}$ generates `back-flow', referring to flow fields generated in response to elastic stresses. 

The active stress has a strength $\zeta$ which is proportional to the density of dipolar sources \cite{Sriram2002}, unique to active systems. In passive systems $\zeta=0$. The sign of $\zeta$ represents the symmetry of the dipolar flow field generated by the active units. In this paper, we only consider $\zeta>0$, corresponding to extensile systems, $\zeta<0$ corresponds to contractile systems. More details about these equations of motion and their application to passive and active systems can be found in \cite{Berisbook, DeGennesBook, Denniston2001, Denniston2004, Davide2007, ourpta2014}. 

The introduction of active stress to the nemato-hydrodynamic equations has non-trivial consequences. The nematic state is unstable to fluctuations of the order parameter field \cite{Sriram2002}. This results in spontaneous flows, which can be stabilised in a 1-D channel \cite{Joanny2005, Davide2007}. In two dimensions the instability leads to active turbulence, shown in Fig.~\ref{fig:actturb}.

\section{Numerical solution and boundary conditions}
\label{sec:numerics}

\begin{figure}
\includegraphics[trim = 0 0 0 0, clip, width=\linewidth]{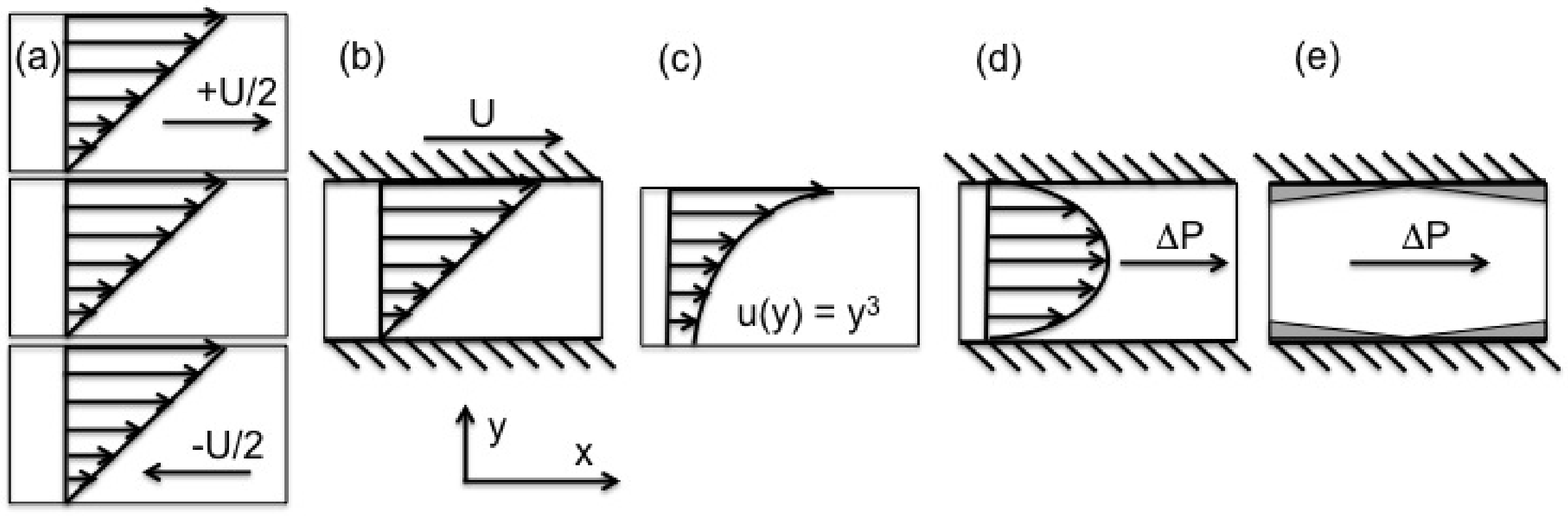}
\caption{Illustrations of various flows and boundary conditions studied in this work. The coordinate system used is also shown. The slanted lines indicate solid boundaries. (a) Lees-Edwards boundary conditions where the domain in the middle is sheared by its own images, (b) Nematic material confined between two parallel plates with the top plate moving with a prescribed velocity, (c) A cubic velocity profile, $u=y^3$ is imposed, (d) A pressure gradient $\Delta P$ is imposed on the nematic material confined between two plates, (e) The channel in panel (d) is made non-uniform with an expanding and contracting cross section.}
\label{fig:flows}
\end{figure}

We use a hybrid lattice Boltzmann method on a $D3Q15$ lattice to numerically solve Eqs.~(\ref{eqn:ns}) and (\ref{eqn:lc}) \cite{Davide2007, Suzanne2011, ourpta2014}. Density variations are considered to be negligible and the numerical scheme essentially yields the solution of the governing equations in the incompressible limit. We study the response of nematic liquid crystals to imposed driving by applying various boundary conditions of increasing complexity:
 
 \begin{enumerate}[i)]
 \setlength\itemsep {1em}
\item \textit{Lees-Edwards boundary conditions:} Typically a material is subjected to shear flow by confining it between two solid plates with one of the plates moving at a prescribed velocity. For nematic liquid crystals, anchoring on the solid boundary may effect the flow and director profiles
and therefore we implement Lees-Edwards boundary conditions, well known in molecular dynamics simulations, to apply a shear flow with no walls (see Fig.~\ref{fig:flows}(a)). We shear the simulation domain by its own image, but with a larger momentum. The director field, $\mathbf{Q}$ undergoes periodic boundary conditions perpendicular to the shearing direction, but with a modified translational velocity. Details of the application of Lees-Edwards boundary conditions to lattice Boltzmann simulations are given in \cite{Wagner2002}.
 
 
\item \textit{Couette flow:} We also subject the nematic material to a shear when confined between two solid plates, with the top plate moving with a prescribed velocity and the bottom plate being stationary (see Fig.~\ref{fig:flows}(b)). By specifying the director orientation on the plates, we can take into account the interaction of nematic material with solid boundaries. We choose homeotropic boundary conditions, with the director field on the boundaries constrained to lie perpendicular to the plates. Comparing the results to those using Lees-Edwards boundary conditions will help us to isolate the effect of boundaries from that of shear. It turns out that this distinction is important in understanding the wall formation.

\item \textit{Nonlinear velocity profile:} We construct a model system to study the response of the director field to an imposed nonlinear flow field. A method analogous to Lees-Edwards boundary conditions which generates a nonlinear velocity profile is not available. Therefore, we subject the nematic material to a prescribed unidirectional flow field with a cubic velocity profile $u = y^3$ where the $y$ direction is normal to the flow (see Fig.~\ref{fig:flows}(c)). We solve only Eq.~(\ref{eqn:lc}) to give the response of the director to this flow profile, neglecting the mass and momentum conservation equations. This model system will help us to distinguish the effect of nonlinear velocity profiles, which are also important in understanding the wall formation process without additional effects arising from solid boundaries. 
 
\item \textit{Poiseuille flow:} A Poiseuille flow profile is produced by confining the nematic material between two solid plates and subjecting it to an imposed pressure gradient (see Fig.~\ref{fig:flows}(d)). As usual in lattice Boltzmann simulations, we impose a body force on the fluid to generate the equivalent effect of a pressure gradient \cite{SucciBook,Guo2002}. No-slip velocity boundary conditions are applied on the bounding plates and homeotropic boundary conditions are specified for the director field. The resulting steady state velocity profile is nonlinear. Thus we have created a natural situation to study the response of the nematic material to nonlinear velocity fields in the presence of solid boundaries, a combination of (ii) and (iii) above. 

\item \textit{Pressure driven flow in channel of non-uniform cross section:} We consider Poiseuille flow in a channel of nonuniform cross section along the imposed flow direction (see Fig.~\ref{fig:flows}(e)). Thus we introduce perturbations to the unidirectional flow field. We consider a channel which is a periodic arrangement of expanding and contracting sections. The half period of the channel is chosen to be 100 lattice points and the expanding and contracting angles are $10^{\circ}$.  As before no-slip/ homeotropic boundary conditions are imposed on the plates for the velocity/director fields.
 
 \end{enumerate}
 
The parameters used in the simulations are $D=0.34$, $A=0.0$, $B=-0.3$, $C=0.3$, $K=0.01$, $\zeta=0.01$, $\mu=2/3$, in lattice units, unless mentioned otherwise. Simulations were performed in a domain of size $100 \times 100$ except in case (v) above where a domain of $200 \times 100$ was chosen. The choice of boundary conditions for the director field on the solid wall, namely homeotropic or planar anchoring, does not affect the conclusions of this work and therefore we chose to report results of simulations using the former alone. The imposed shear rate  (with or without solid boundaries) is $10^{-4}$. The applied pressure gradient generating Poiseuille flow in the channel geometries is $5 \times 10^{-5}$. As usual in lattice Boltzmann schemes, discrete space and time steps are chosen as unity and therefore all quantities can be converted to physical units in a way which depends on the material of interest \cite{Cates2008, Henrich2010, ourprl2013}. This procedure will demand choosing appropriate mass, length and time scales. Molecular liquid crystals are typically $\sim 10$nm in length, exhibit a rotational diffusivity $\sim$ 100-1000 (Pa-s)$^{-1}$ and have elastic constants $\sim$ 1pN \cite{DeGennesBook}; and these can be used as an appropriate basis for conversion to physical units. However, for active liquid crystals, similar data are not available yet. A similar system may be FD viruses, passive colloidal liquid crystals, which are microns in length, have rotational diffusivity $\sim$ 20 (Pa-s)$^{-1}$ and elastic constants $\sim$ 1pN.
 
 \section{Driven passive nematics}
 \label{sec:passive}
 
\begin{figure}
\center
 \subfigure[]{\includegraphics[trim = 60 30 50 20, clip, width=0.37\linewidth]{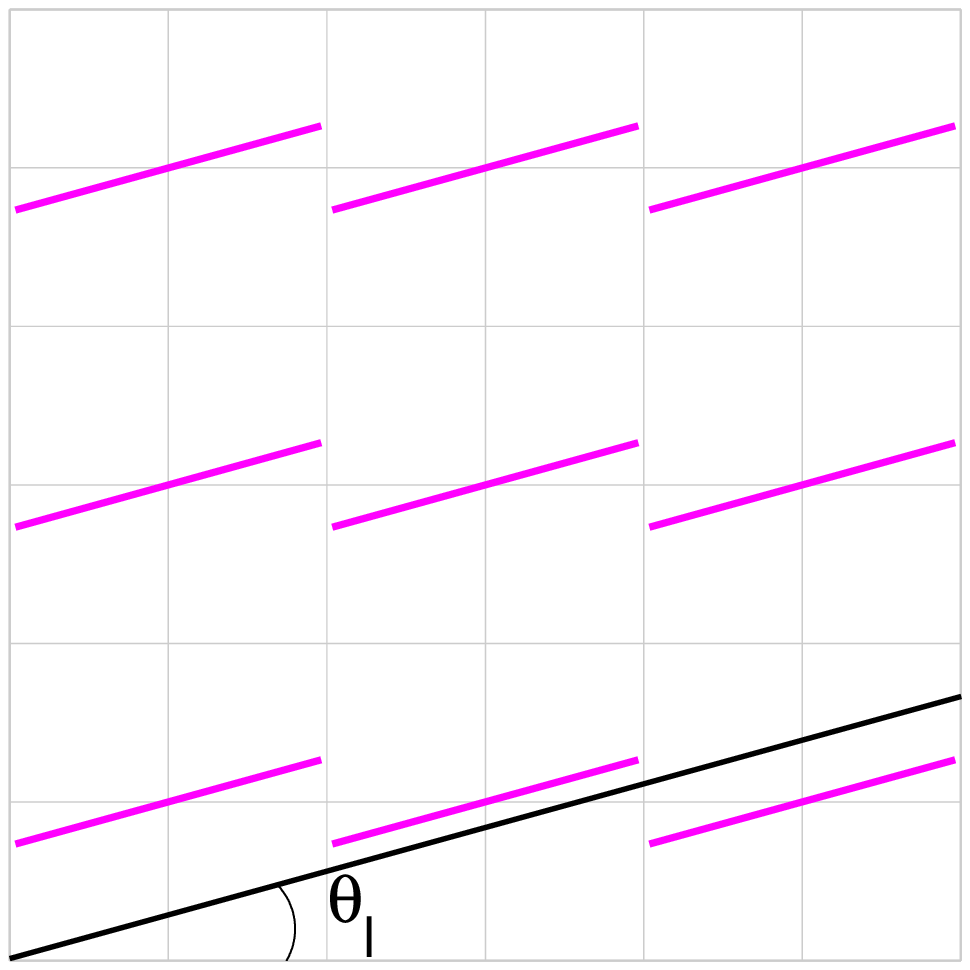}\label{fig:leeedalign}}
  \subfigure[]{\includegraphics[trim = 60 30 25 20, clip, width=0.4\linewidth]{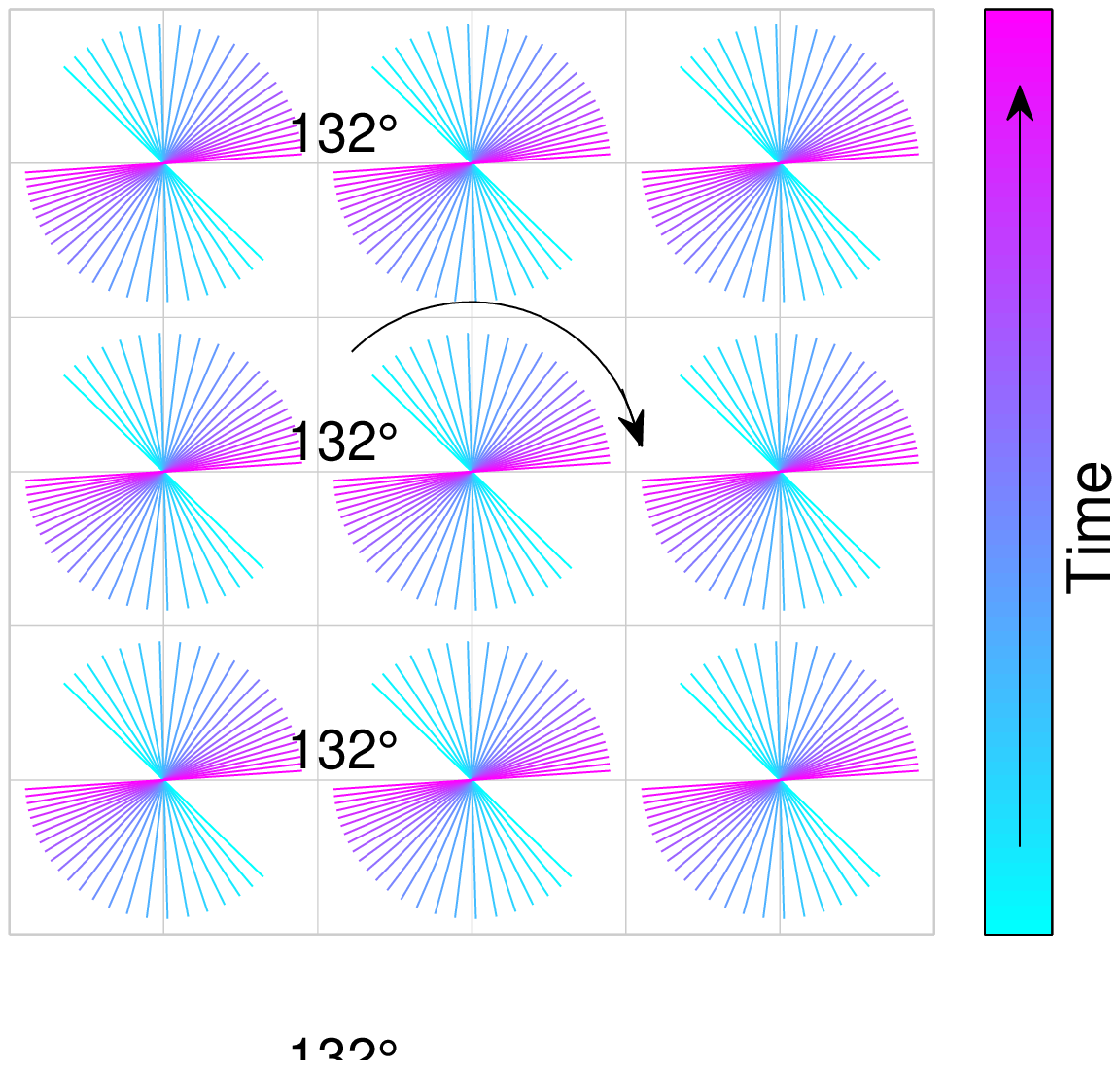}\label{fig:leeedtumbl}}
  \caption{Response of the nematic director field to a simple shear flow imposed by Lees-Edwards boundary conditions (see Fig.~\ref{fig:flows}(a)). The director field (only $9$ grid points are shown here), which rotates in the clockwise direction following the vorticity, is shown at equal intervals of time, colour-coded blue to magenta (see text for a detailed explanation). (a) $\lambda_1>1$: the director field aligns at the Leslie angle. (b) $\lambda_1<1$: the director tumbles. The angular velocity is largest when the director is along the flow gradient direction.}
  \label{fig:leeed}
 \end{figure}
 
We first summarise the response of a passive, two-dimensional, nematic suspension to a simple shear flow imposed using Lees-Edwards boundary conditions.  
Fig.~\ref{fig:leeed} compares the behaviour of aligning and tumbling materials. For $\lambda_1 > 1$ the director aligns at the Leslie angle, $\theta_l = \frac{1}{2}\cos^{-1}({1}/{\lambda_1})$, with respect to the flow direction  as shown in Fig.~\ref{fig:leeedalign}. 
For $\lambda_1 < 1$ the director undergoes continuous tumbling, and  Fig.~\ref{fig:leeedtumbl} shows its time evolution during part of a rotation period. This, and subsequent, figures should be read as follows: The time has been given a color coding represented by the colorbar on the right hand side. The orientation of the director at the initial time is shown in blue. As time proceeds, the director rotates following the imposed vorticity. This rotation is in the clockwise direction as indicated by the curved arrow. The final orientation of the director is shown in magenta. Intermediate orientations at equally spaced time intervals, which correspond to the color-bar, are also shown. Just as for a single particle in a simple shear flow, the maximum angular velocity occurs when the director field is oriented along the gradient direction ($y$-axis) and the minimum when it is in the direction of the flow ($x$-axis). We also give the difference in angle between the chosen final and initial time - this will be relevant when the number of rotations varies with position.

So far we have observed that nematics subjected to a simple shear without any solid boundaries and confined to two dimensions remain homogeneous in space. There are no walls or topological defects that are characteristic features of active turbulence. We now investigate the essential ingredients that are required for walls and topological defects to form in passive nematics.
 
 \subsection{Walls in the director field}
\begin{figure}
\center
 \subfigure[]{\includegraphics[trim = 80 30 120 20, clip, width=0.4\linewidth]{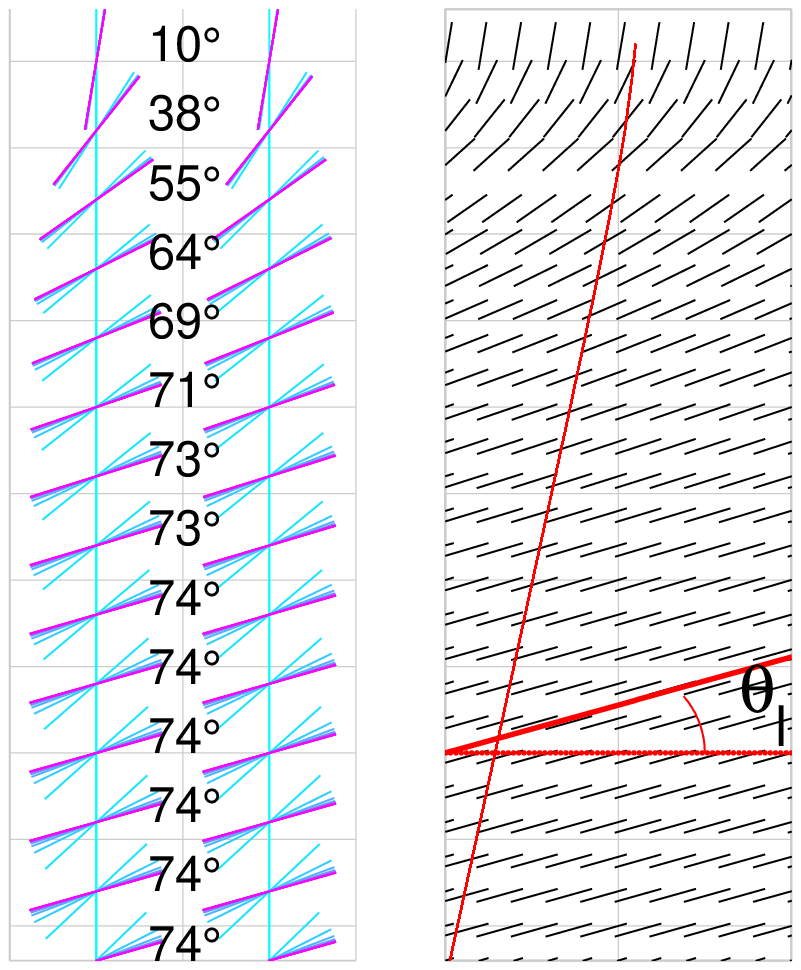}\label{fig:couettealign}}
  \subfigure[]{\includegraphics[trim = 80 30 110 20, clip, width=0.45\linewidth]{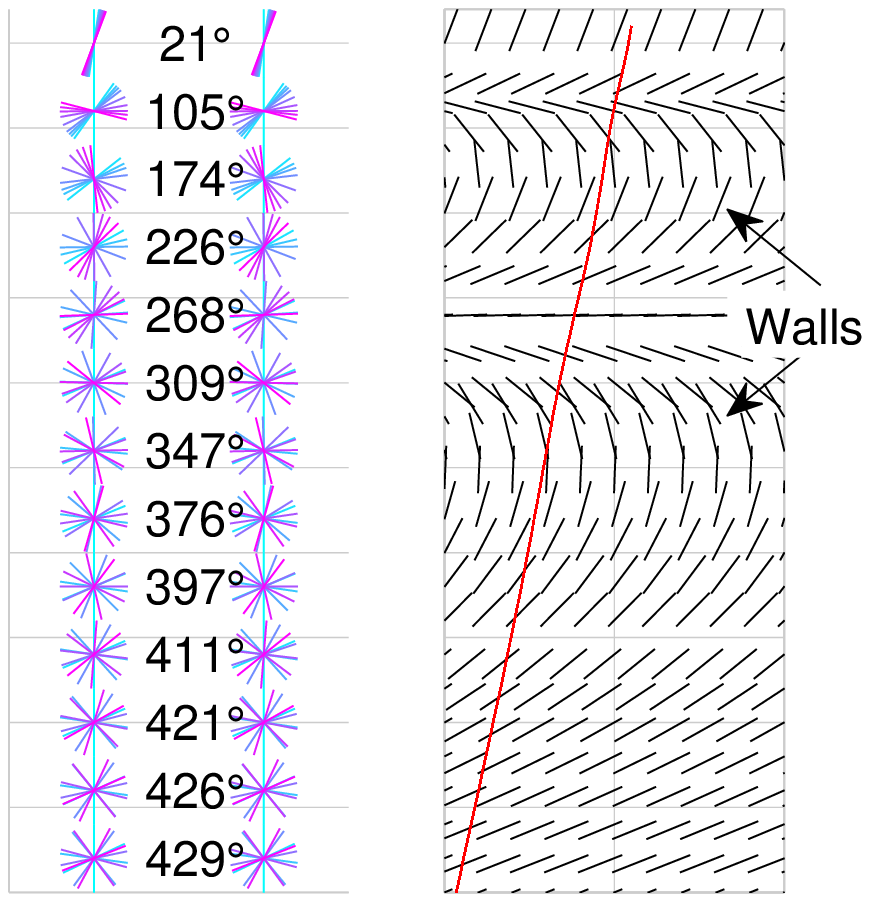}\label{fig:couttetumbl}}
  \caption{Wall formation in the director field when an (a) flow aligning and (b) flow tumbling nematic is subjected to Couette flow. Only the top half of the domains are shown. The illustrations on the LHS of panels (a) and (b) follow the conventions of Fig.~\ref{fig:leeed}. The RH plot is the director field corresponding to the last frame of the LH plot. Imposed homeotropic boundary conditions on the plates do not allow spatially homogeneous tumbling of the director field. Thus, in the flow aligning case (a), a half wall is formed near the top solid plate and the director aligns at the Leslie angle everywhere else. In (b) the director field tumbles, and spatial inhomogeneities generate walls. The flow profiles in each case are also shown.
  }
  \label{fig:couette}
 \end{figure}
  
\begin{figure}
\center
\subfigure[]{\includegraphics[trim = 80 30 120 20, clip, width=0.4\linewidth]{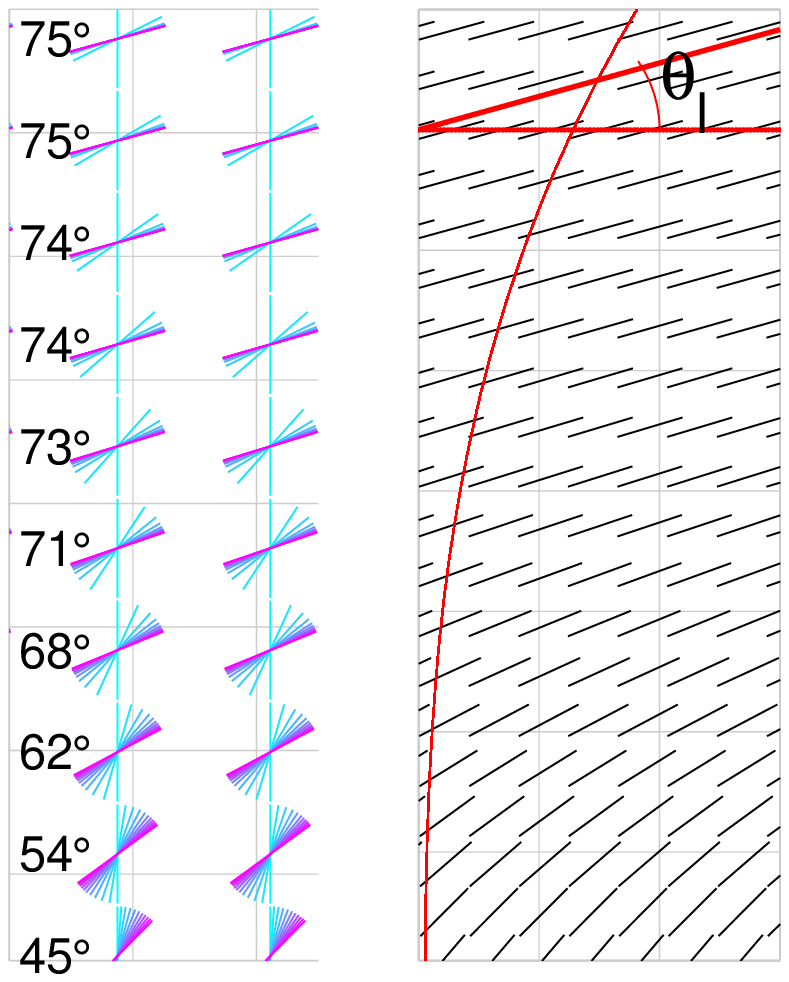}\label{fig:nlinshearalign}}
\subfigure[]{\includegraphics[trim = 80 30 120 20, clip, width=0.4\linewidth]{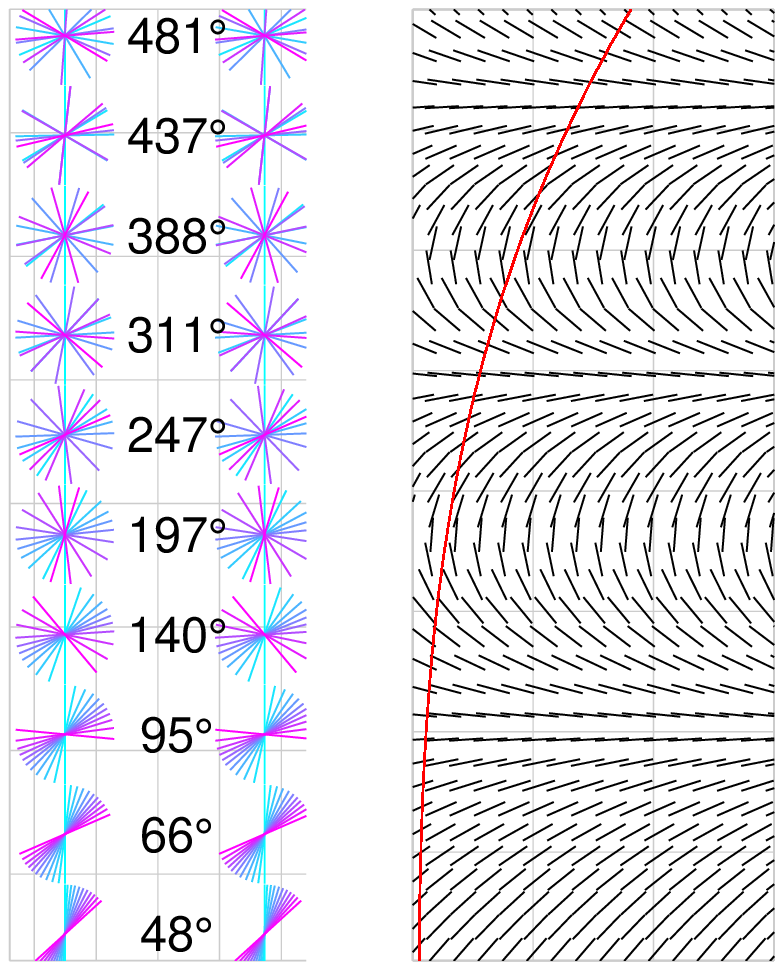}\label{fig:nlinsheartumbl}}
\caption{Wall formation in the director field when an (a) flow aligning and (b) flow tumbling nematic is subjected to a cubic velocity profile. The illustrations on the LHS of panels (a) and (b) follow the conventions of Fig.~\ref{fig:leeed}. The RH plot is the director field corresponding to the last frame of the LH plot. The spatial gradient in vorticity causes the director field to rotate with different angular velocities giving (a) an inhomogeneous orientation field in the aligning case and (b) walls in a tumbling material. The flow profiles in each case are also shown.}
\label{fig:nlinshear}
 \end{figure}

Both walls and topological defects are inhomogeneous director fields. 
Therefore we now investigate whether inhomogeneous forcing can generate wall-like structures. Indeed we find that walls naturally form if the angular velocity of the director field in response to flow gradients or elastic forces is inhomogeneous in space.
We show that there are two different ways to impose director inhomogeneities (i) via the introduction of solid boundaries and (ii)  by imposing nonlinear flow fields.

We first consider Couette flow with homeotropic anchoring specified on both bounding plates. The simulations were started with director field normal to the plates, consistent with the boundary conditions. The top plate is allowed to move and the resulting configurations of the director field are illustrated in Fig.~\ref{fig:couette} for both a shear aligning and a shear tumbling material.  Only the top half of the channel is shown, as the configurations have reflection symmetry about the centre of the channel. For flow aligning nematics, Fig.~\ref{fig:couette}(a) shows that the bulk of the sample is rotated by approximately $74^{\circ}$ (in this case) and aligned with the flow direction at the Leslie angle ($\approx 16^{\circ}$). This is the same behaviour as for Lees-Edwards boundary conditions. However, near the solid boundary, the homeotropic anchoring competes with the flow aligning tendency of the director field, thus generating a smooth variation in the orientation. The result is the formation of a half wall near each boundary.  

In the case of tumbling nematics in Couette flow there is a similar competition between the anchoring boundary conditions and the rotation of the director field as shown in Fig.~\ref{fig:couette}(b). Unlike the aligning suspensions, the response of tumbling materials is not uniform throughout the bulk. This is because, as time proceeds, the director field continues to rotate instead of approaching a constant angle of orientation and the total angle rotated  increases between the edges and the centre of the channel (from approximately $21^{\circ}$ to $429^{\circ}$ in this case). This spatially differential tumbling results in walls in the director field as shown in Fig.~\ref{fig:couette}(b). We stress that Fig.~\ref{fig:couette}(b) is not a steady state, tumbling of the director field continues in response to the imposed flow and flow gradients. Thus a dynamic situation persists with continual formation and destruction of walls in the director field.


Secondly, we show that differential rotation rates can also be generated by imposing nonlinear velocity fields. We solve Eq.~(\ref{eqn:lc}) with an
imposed cubic velocity profile. As before the simulations were started with an initial director field oriented along the $y$ direction. The results are illustrated in Fig.~\ref{fig:nlinshear}.

 In the case of flow aligning nematics, the director field rotates towards the Leslie angle. Note that the Leslie angle is independent of the shear rate and depends only on the value of $\lambda_1$. The rotation is fastest where the velocity gradient is largest (near top of the domain). The bottom of the domain experiences smaller shear and had not rotated enough to reach the Leslie angle during this simulation. 
In the case of tumbling nematics, the director field continuously rotates. Due to the cubic velocity profile, vorticity is not constant in space and hence different regions along the $y$ direction rotate with different angular velocities. While the director at the top of the domain has rotated more than a period, the bottom has rotated only by a quarter of a period. The result is the formation of walls. Again, Fig.~\ref{fig:nlinshear}(b) is not a steady state picture, but a representation of a temporally evolving state.

 \begin{figure}
 \center
 \subfigure[]{\includegraphics[trim = 80 30 120 20, clip, width=0.4\linewidth]{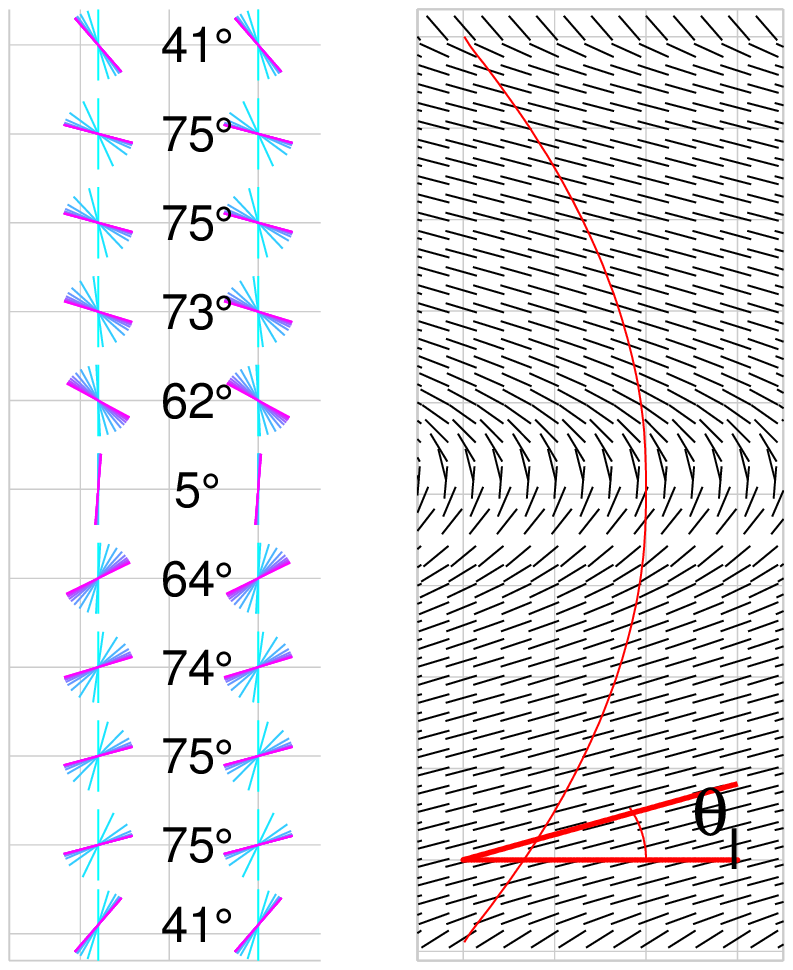}\label{fig:poiseualign}}
\subfigure[]{\includegraphics[trim = 80 30 120 20, clip, width=0.4\linewidth]{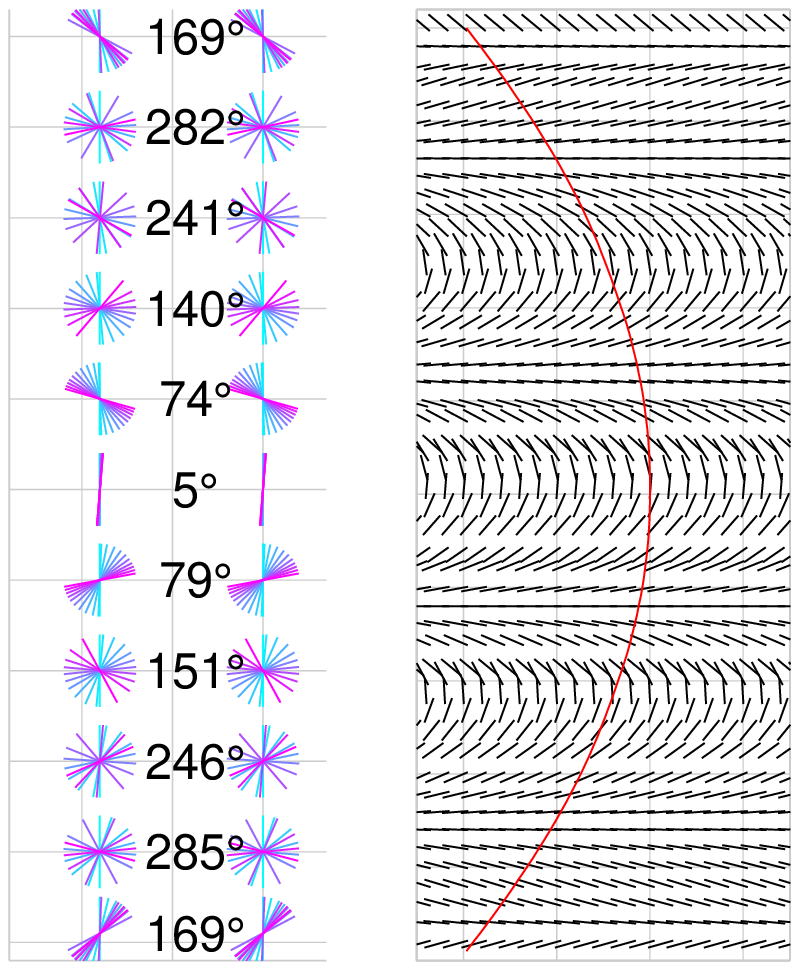}\label{fig:poiseutumbl}}
\caption{Wall formation in the Poiseuille flow of nematic liquid crystals. 
 The illustrations on the LHS of panels (a) and (b) follow the conventions of Fig.~\ref{fig:leeed}. The RH plot is the director field corresponding to the last frame of the the LH plot.
(a) Aligning suspensions: the solid boundaries generate half walls and flow gradients generate a wall at the centre of the channel. (b) Tumbling suspensions: the solid boundaries and nonlinear velocity profile act together to generate walls. The flow profiles in each case are also shown.}
\label{fig:poiseu}
 \end{figure}
 
Thus the differential rotation of the director field in response to vorticity gradients, caused by boundaries or varying shear, generates inhomogeneities in the director field which, in tumbling suspensions, may be identified as walls. In a Poiseuille flow geometry both boundaries and shear gradients are present. Therefore we now consider the flow of a nematic material between two solid plates, with homeotropic boundary conditions on the director, under a pressure gradient. 
The results are shown in Fig.~\ref{fig:poiseu}. The simulations were started with an initial director field in the vertical direction. 

For aligning nematics, the director field rotates and reaches a steady state at the Leslie angle everywhere except near the boundaries and in the centre of the channel. As for Couette flow, the competition between the flow alignment and the boundary anchoring leads to the formation of half-wall structures near both boundaries. In the centre of the channel, where the slope of the velocity profile changes sign, the director field rotates to generate a wall. As the driving pressure gradient increases, the bend configuration of this wall (known as a `bow' mode in the literature) is replaced by a `peak' configuration \cite{Denniston2001b}. Such changes in director configuration are known to change the rheology of the material  \cite{PasechnikBook}. Flow of an aligning nematic under an imposed pressure gradient with various anchoring conditions has been recently investigated in \cite{Matthewarxiv} due to its importance in microfluidics. This study also reports similar director configurations. In the case of pressure driven flow in three dimensional channels, the director field need not be constrained to two dimensional planes and walls can be replaced by three dimensional director configurations arising from a competition between flow and anchoring effects \cite{Anupam2013b}. Small differences in anchoring are known to lead to large differences in the director configurations \cite{Sambles2010}. 

In case of flow tumbling nematics under pressure driven flow we observe several walls generated by the boundary anchoring and the nonlinear velocity profile acting together. Again, the director field does not reach a steady state, but instead walls are continually created, destroyed and displaced. The corresponding velocity profiles, shown for both aligning and tumbling nematics in Fig.~\ref{fig:poiseu}, have no noticeable structure at the location of walls.

The complex behaviour arising from the competition between flow and ordering effects in passive nematic liquid crystals and colloidal rods have been experimentally demonstrated in some recent investigations \cite{Lettinga2005, Anupam2014, Anupam2011, Anupam2013}. Our simulations are distinct from these experiments and most previous work in that we constrain the director field to two dimensions and do not allow the director to escape in the vorticity direction on the application of a shear force. This is to give a direct comparison with experimental active systems 
such as actin or microtubule filaments driven by molecular motors \cite{Dogic2012} and bacterial suspensions \cite{Sokolov2007} where the active nematic material was constrained in two dimensional films. Moreover, in these cases, the length characterising the constituents is on much larger scales than typical molecular liquid crystals making escape into the third dimension less likely.

\begin{figure}
\subfigure[]{\includegraphics[trim = 40 80 40 70, clip, width=0.49\linewidth]{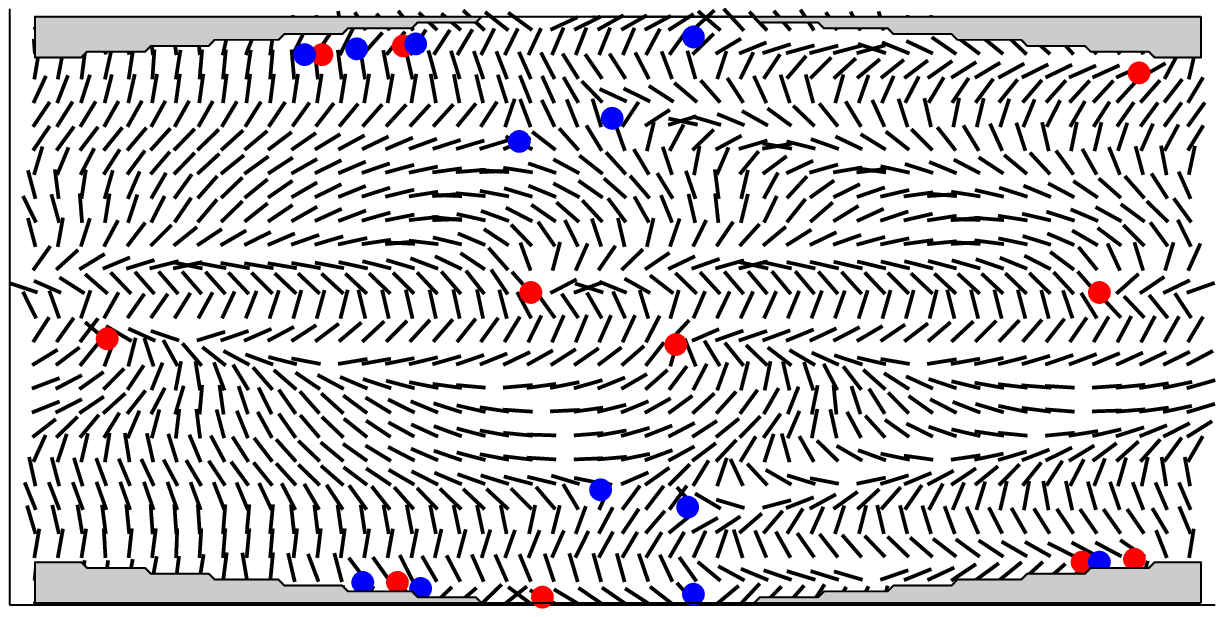}\label{fig:wedgepassdir}}
\subfigure[]{\includegraphics[trim = 40 80 40 70, clip, width=0.49\linewidth]{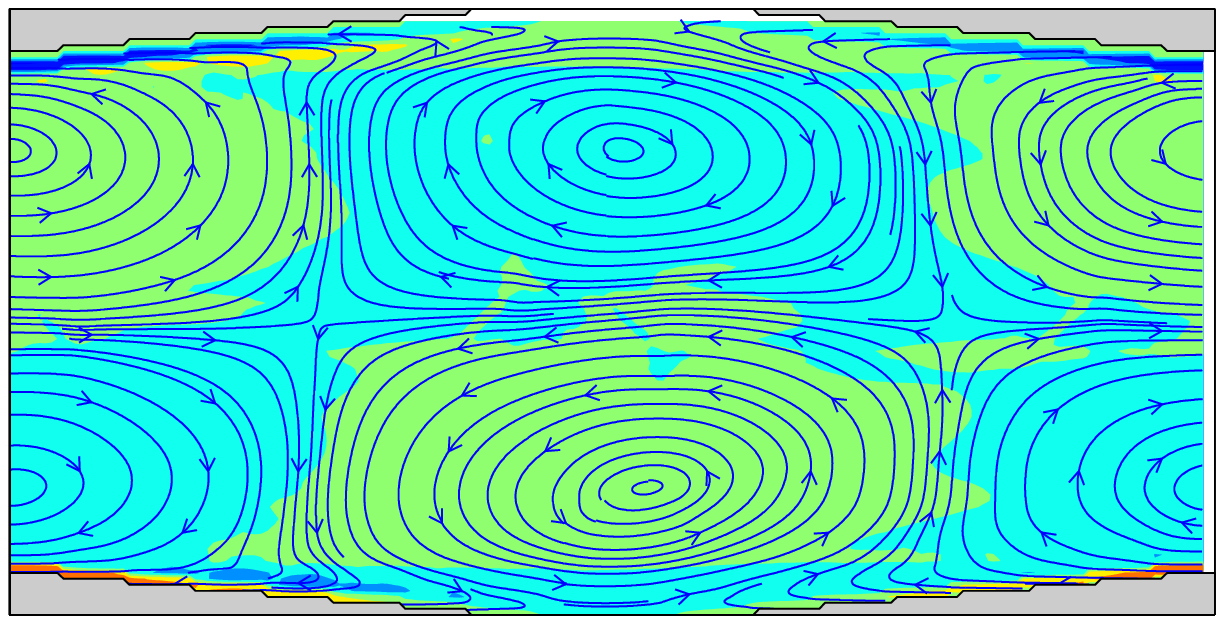}\label{fig:wedgepassvel}}
\caption{(a) Tumbling passive nematics subject to pressure-driven flow  in a channel of non-uniform cross section generate walls and defects. (b) corresponding flow profile with the average velocity in $x-$direction subtracted. Both streamlines and vorticity field (colour shading) are shown.}
\label{fig:wedge}
\end{figure}

\subsection{Defect formation from walls}
So far, we have been looking at the formation of walls in nematics. For all the different situations that we have studied, 
 walls did not show any instabilities leading to defect formation as observed in active turbulence (see Fig.~\ref{fig:actturb}(b)). This is a consequence of the one dimensional nature of the imposed forces. Walls can become unstable if there are sufficiently large perturbations or inhomogeneities. We introduce the latter by using a channel of non-uniform cross section as shown in Fig.~\ref{fig:wedge}. We consider tumbling nematics in Poiseuille flow with homeotropic anchoring at the solid boundaries. As discussed in the previous section walls are generated due to shear. However, now the walls are slightly perturbed along the flow direction due to variations in the flow and director field in the expanding-contracting cross section of the channel. Such perturbations are sufficient to drive the walls unstable and they break down to form a pair of oppositely charged topological defects. The defects both destroy the walls and regenerate nematic orientational order. Due to the imposed flow, they are carried away and execute their own motion driven by elastic forces and back flow effects,  annihilating on finding defects of opposite charge. Thus, walls and defects are continuously created and destroyed in a tumbling nematic suspension when driven in a channel of non-uniform cross section. 

This can be considered the counterpart of active nematics where these phenomena also occur but driven by active forces. However, unlike for active turbulence the presence of walls and defects in passive tumbling nematics does not alter the flow field substantially. Panel (b) of Fig.~\ref{fig:wedge} shows the vorticity and streamlines of the flow field once the average horizontal velocity is subtracted. The circulating flows are a consequence of mass conservation in the expanding-contracting cross section of the channel and are not related to walls or defects.
We also note that walls can disintegrate into topological defects even in a channel of uniform cross section if thermal fluctuations are sufficiently large. 


\section{Driven active nematics}
\label{sec:active}

\subsection{Suppression of hydrodynamics instabilities under shear}

\begin{figure}
\centering
\subfigure[$\zeta=0.002$]{\includegraphics[trim = 0 0 0 0, clip, width=0.4\linewidth]{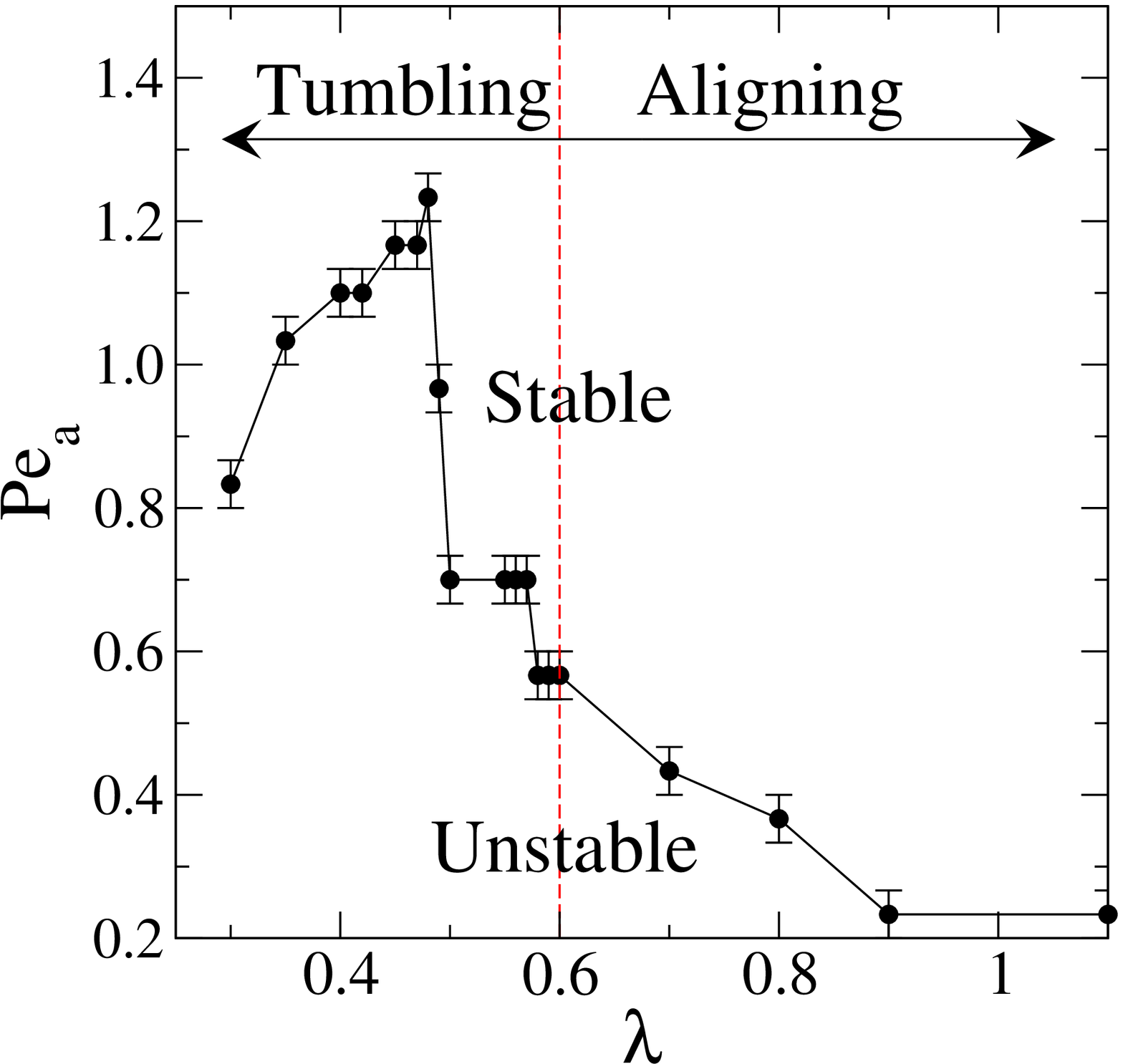}\label{fig:shearactive}}
\subfigure[$\zeta=0.002, \lambda=0.4$]{\includegraphics[trim = 90 35 75 25, clip, width=0.4\linewidth]{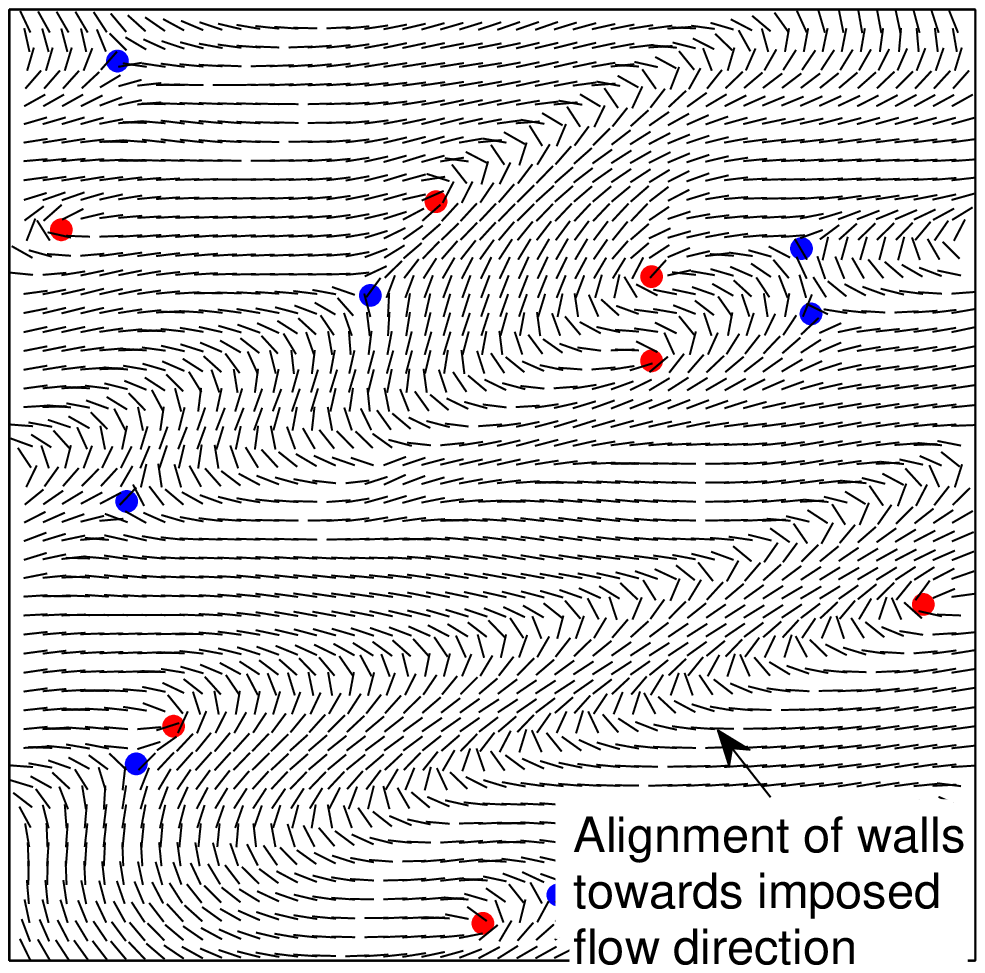}\label{fig:wallalign}}\\
\subfigure[$\zeta=0.003$]{\includegraphics[trim = 90 35 75 25, clip, width=0.4\linewidth]{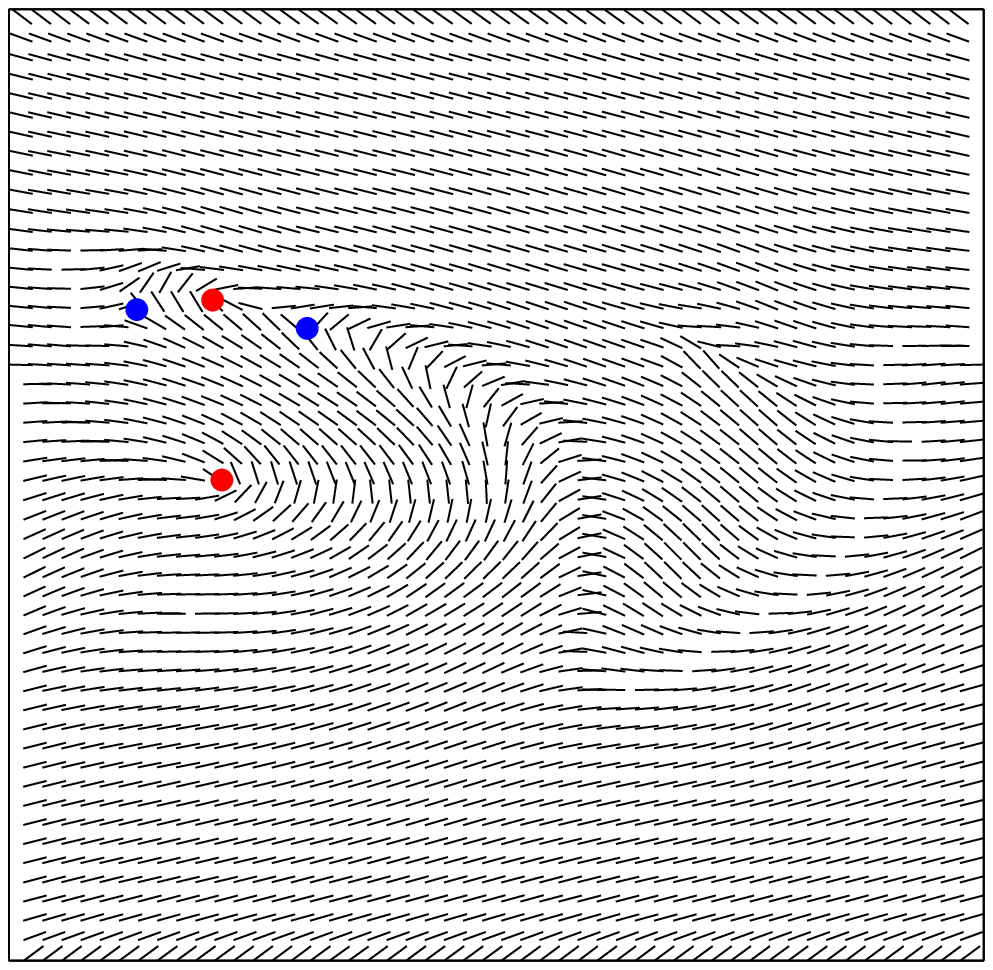}\label{fig:activepois}}
\subfigure[$\zeta=0.012$]{\includegraphics[trim = 90 35 75 25, clip, width=0.4\linewidth]{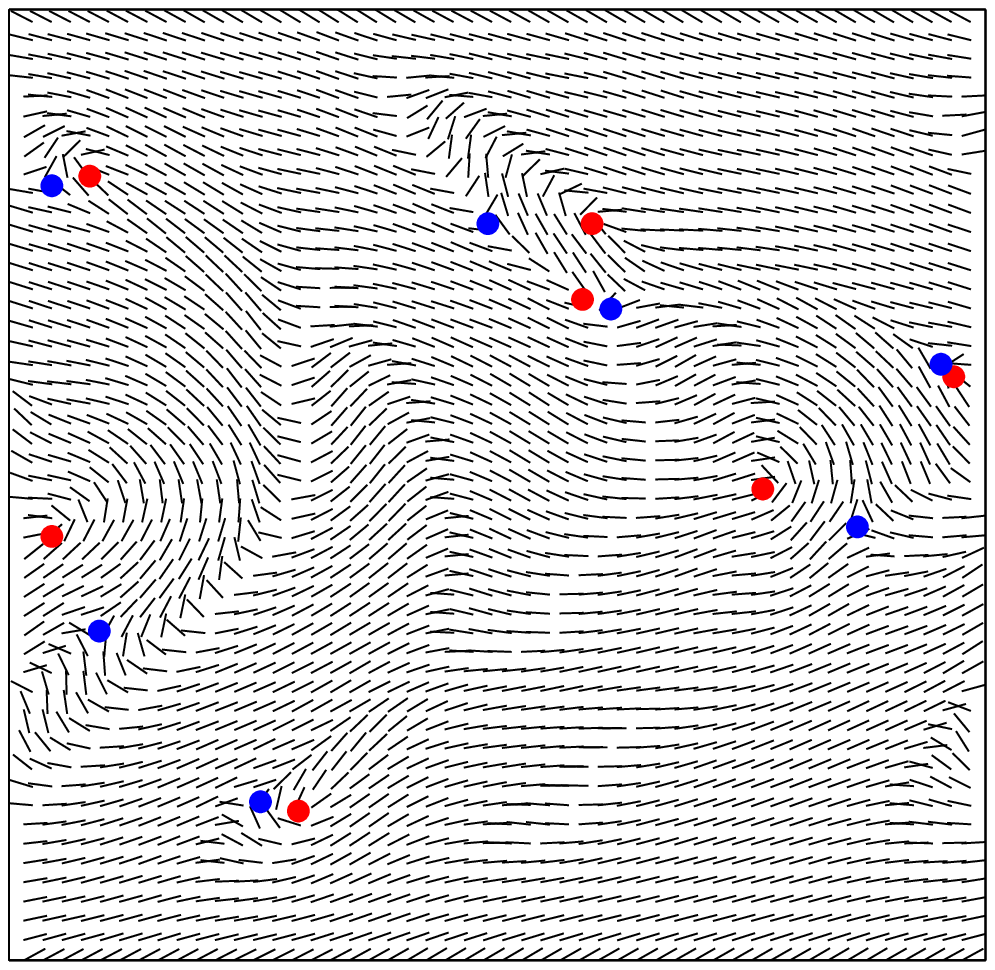}\label{fig:actpoiselarge}}
\caption{Suppression of the hydrodynamic instabilities of an active nematic  by shear. 
(a) phase diagram showing the stable and unstable regions as a function of $Pe_a$, the ratio of imposed driving to active driving, and $\lambda$, the alignment parameter. Simulations were performed in a domain of size $100 \times 200$. Both aligning and tumbling regimes are shown. (b) alignment of walls due to simple shear before the hydrodynamic instabilities are completely suppressed. (c) active turbulence in Poiseuille flow. Instabilities and thus active turbulence are seen only at the centre of the channel where the shear rate due to the imposed pressure gradient is weak. (d) The width of the turbulent layer increases as activity increases.}
\label{fig:supp}
\end{figure}

We now consider driven active nematics, thus studying situations in which both active and external forces are acting together.
Firstly, active nematics are subjected to simple shear flow by applying Lees-Edwards boundary conditions.  This situation was studied by Muhuri {\it et al.} \cite{Muhuri2007} who used linear stability analysis to show that hydrodynamic instabilities are suppressed by shear and that the critical shear rate needed to stabilise the active nematic increases with decreasing aligning parameter, $\lambda$.
They obtained a phase diagram of $Pe_a$ vs $\lambda$ where $Pe_a = 2 \mu \gamma / \zeta$ measures the ratio of the imposed driving to the active driving, with $\gamma$ the imposed shear rate. We construct a similar plot, Fig.~\ref{fig:shearactive}, by numerically solving the entire set of nonlinear governing equations, confirming their analysis beyond the linear regime. 
Moreover our simulations show that as $\lambda$ decreases into the tumbling regime the critical shear rate for stability continues to increase, consistent with the fact that tumbling materials have an intrinsic tendency to generate walls. However, simulations show that when $\lambda \approx 0.5$ the critical shear rate decreases again. The reasons for this change in behaviour are not yet clear.

Active nematics exhibit an interesting behaviour at shear rates close to the value required for stabilising the nematic. The walls generated by the active stresses themselves respond to the imposed shear field. Behaving effectively as a single structure, they align close to the flow direction.
 This wall alignment regime is illustrated in Fig.~\ref{fig:wallalign}. Defects may continue to be produced, so that the walls continually disappear, and then re-form.

 \subsection{Poiseuille flow}
 
The competition between the stabilising effect of shear and the destabilising effect of the activity is also seen in an active nematic under Poiseuille flow. The imposed pressure gradient generates a velocity profile with a maximum at the centre of the channel as in Fig.~\ref{fig:poiseu}. Shear rates are smallest at the centre of the channel and thus active stress can dominate the imposed shear rate and walls and defects form. Nearer the walls the shear rates are sufficiently large that active hydrodynamic instabilities are suppressed and  the nematics align at the Leslie angle. Hence active turbulence is confined to the centre of the channel as illustrated in Fig.~\ref{fig:activepois}. If activity is further increased the width of the turbulent layer in the channel is increased. At sufficiently large activities, activity driven flow will dominate the imposed flow. 

\begin{figure}
\begin{minipage}{0.6\linewidth}
\subfigure[]{\includegraphics[trim = 40 80 40 70, clip, width=0.9\linewidth]{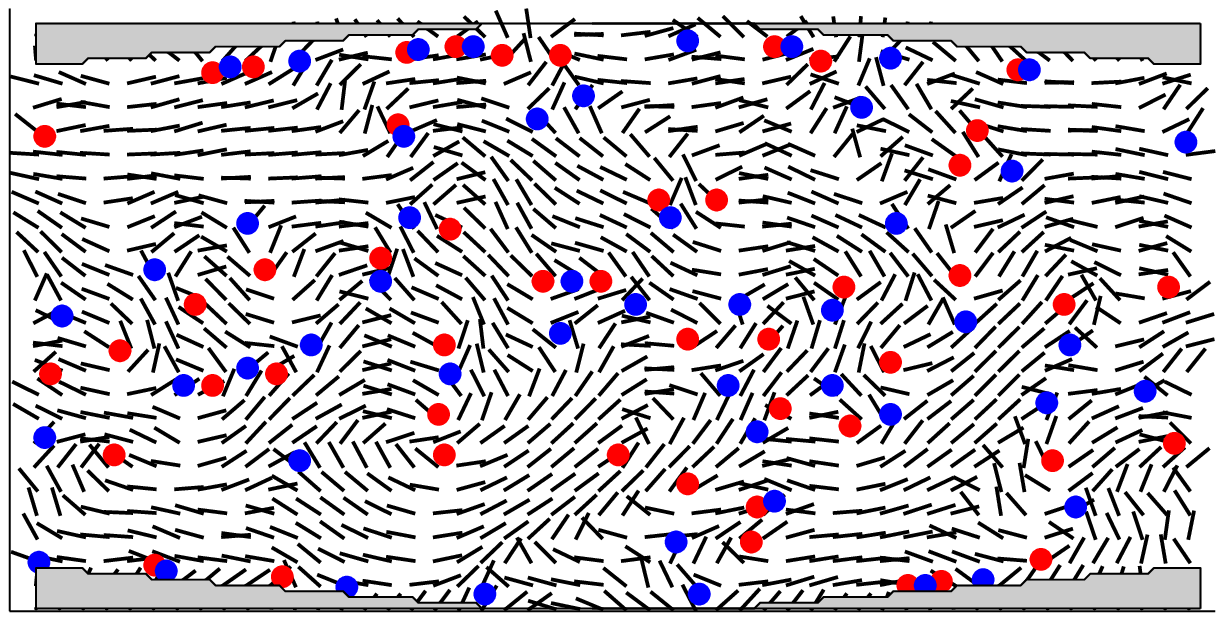}\label{fig:wedgeactdir}}
\subfigure[]{\includegraphics[trim = 40 80 40 70, clip, width=0.9\linewidth]{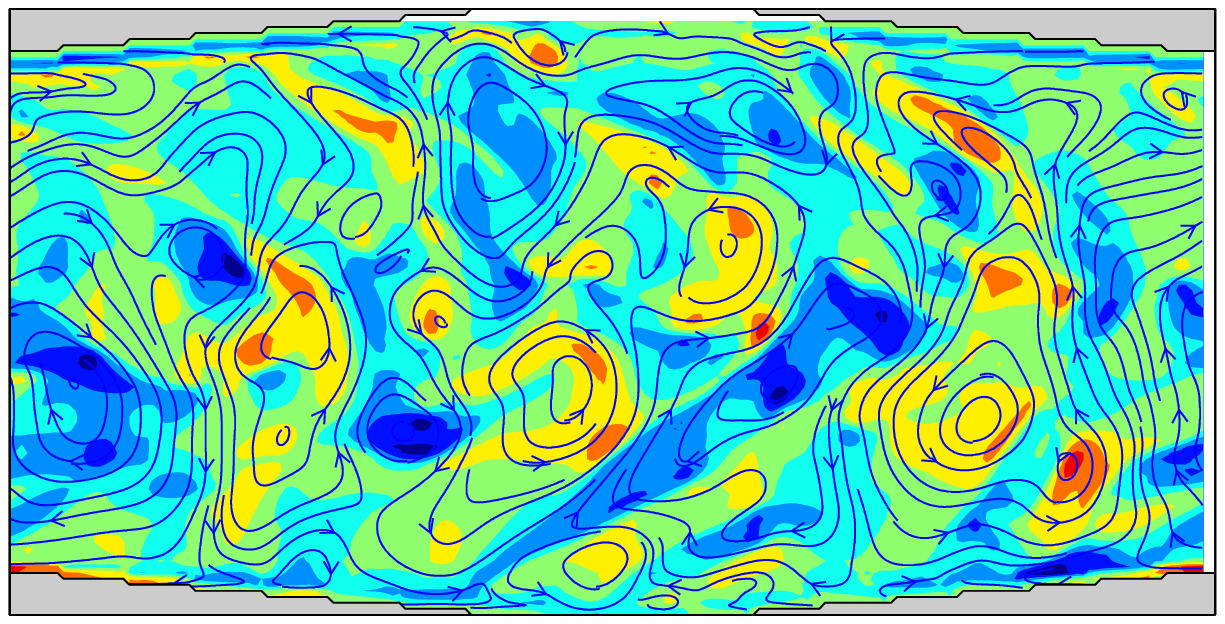}\label{fig:wedgeactvel}}
\end{minipage}
\begin{minipage}{0.35\linewidth}
\subfigure[]{ \includegraphics[trim = 0 0 0 0, clip, width=\linewidth]{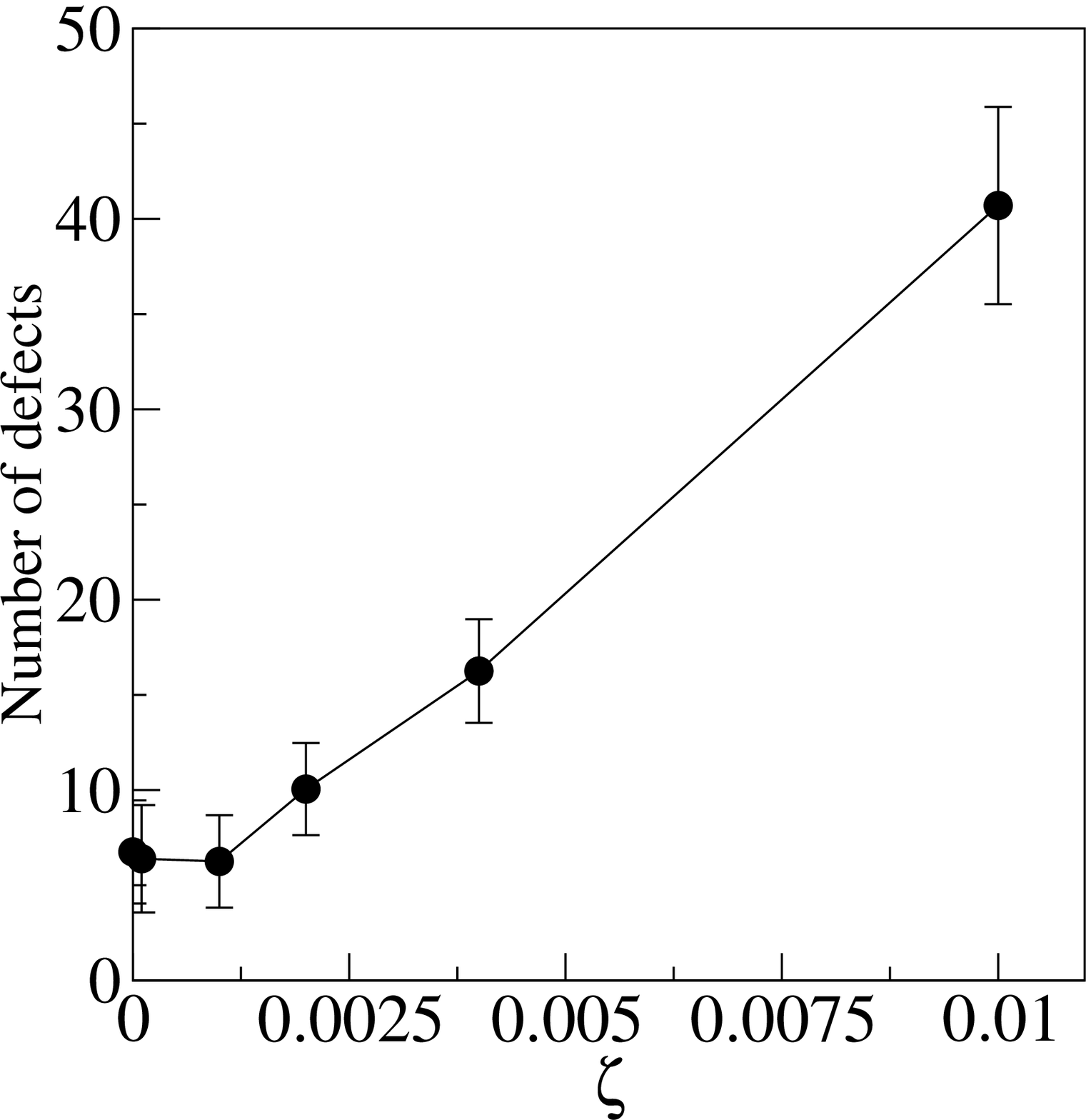}\label{fig:ndefects}}
\end{minipage}
\caption{Tumbling active nematics driven by an imposed pressure gradient in a non-uniform channel. (a)  walls and defects.  (b) flow field with the average velocity in the $x-$direction subtracted. Both streamlines and vorticity (colour-shading) are shown. (c) Number of defects as a function of activity $\zeta$. The number of defects tends to a constant value as the activity is reduced.}
\label{fig:activewedge}
\end{figure}

To complete the picture and to compare to passive nematics we now consider the behaviour of active tumbling nematics in a periodically expanding-contracting channel driven by a pressure gradient.  The results for active nematics are illustrated in Fig.~\ref{fig:activewedge} (compare
Fig.~\ref{fig:wedge} for the passive case).  Panel (a) shows the director field, dominated by walls and defects as both the active stress and the imposed pressure gradient generate these director textures. Panel (b) shows the corresponding vorticity and stream lines, again after subtracting the mean flow along the channel. This is very different to the passive case as the defects produce additional flows through the active term in the stress tensor.

To link the active and passive systems we plot the number of defects at different activity levels in Fig.~\ref{fig:ndefects}. 
As activity decreases, so the number of defects reduces. In the case of a non-driven system, the number of defects would reduce to zero at sufficiently small activities \cite{ourprl2013}. However, here, a finite number of defects is generated even at zero activity due to the driving alone.

%
  
 \section{Discussion}
 \label{sec:discussion}
 
  \begin{figure}
 \subfigure[]{\includegraphics[trim =  0 0 0 0, clip, width=0.49\linewidth]{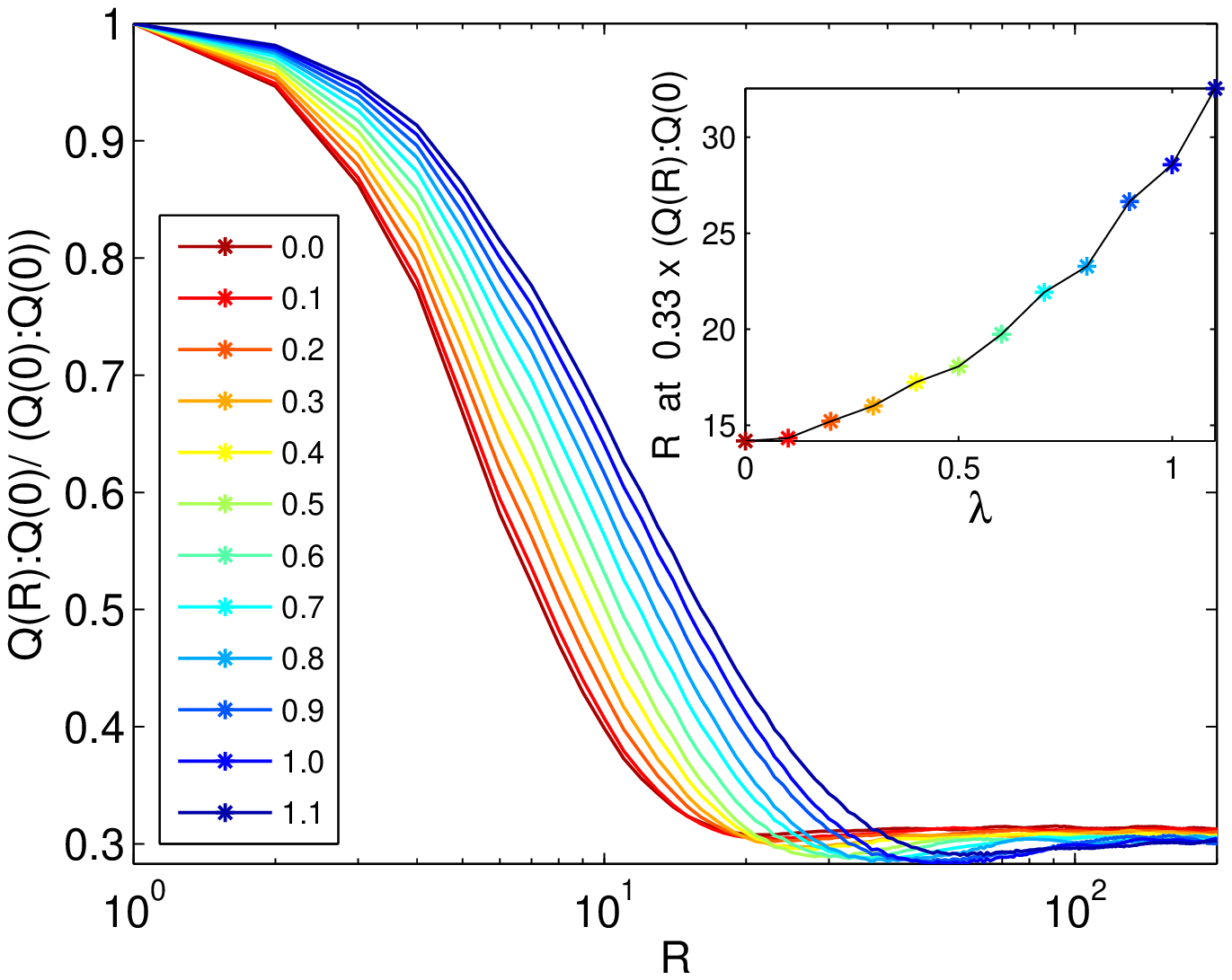}\label{fig:nnext}}
 \subfigure[]{\includegraphics[trim = 0 0 0 0, clip, width=0.49\linewidth]{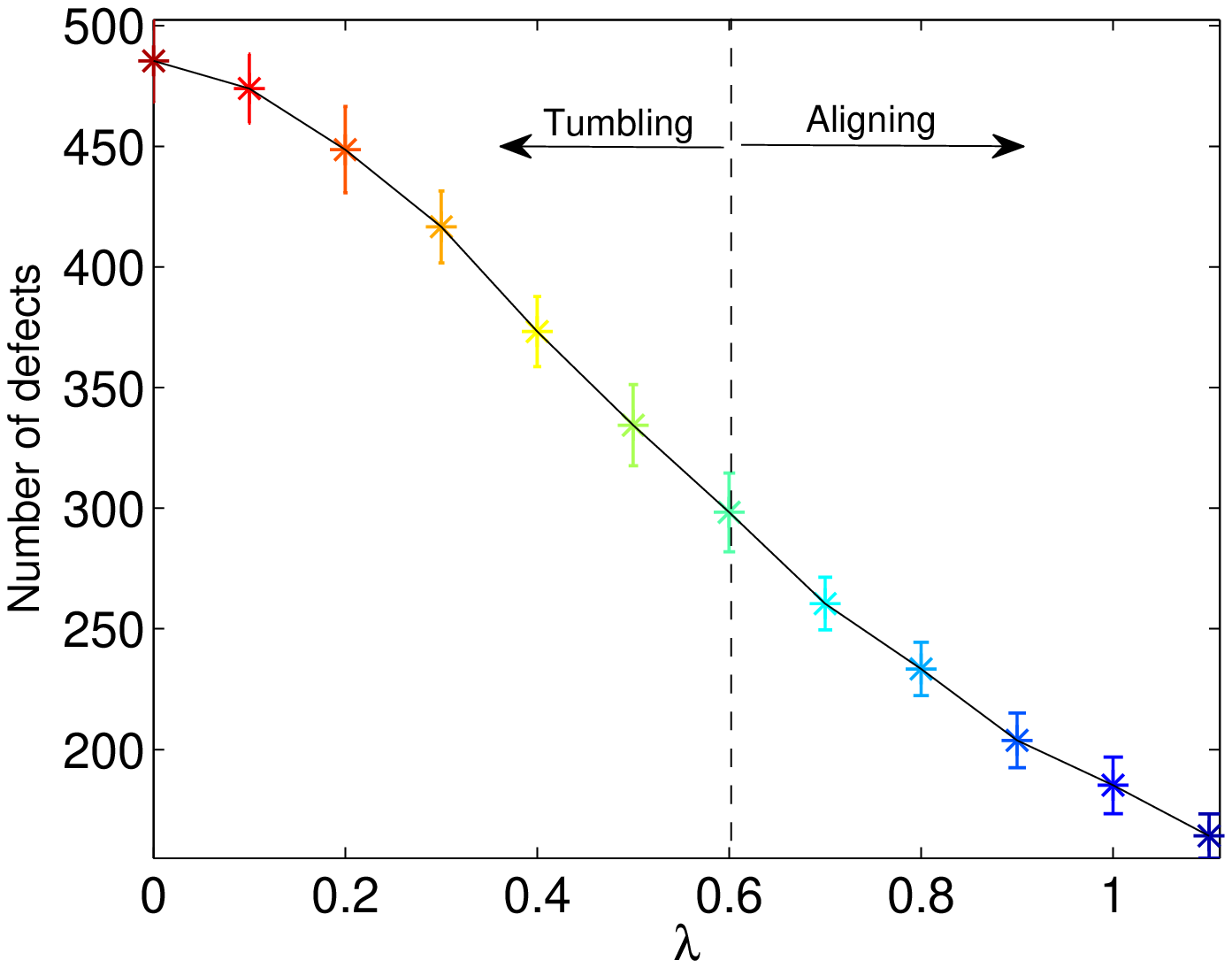}\label{fig:ndeflambda}}
\caption{(a) Spatial correlations of the nematic director field for various values of the alignment parameter $\lambda$ for an active nematic ($\zeta=0.01$). Both aligning and tumbling regimes are shown.} The characteristic length of the nematic order decreases as $\lambda$ decreases due to the appearance of more walls and topological defects. This is also apparent from  (b) the number of defects as a function of the alignment parameter $\lambda$.
\label{fig:lambdaeffects}
 \end{figure} 

Our simulations were in 2D with the director field confined to the plane to give a direct comparison to most previous work on active systems. 
Therefore we see flow aligning and tumbling regimes under shear but not the instabilities that lead to kayaking, log rolling or rheochaos which require escape of the director into the third dimension, or wagging which would only appear at much higher shear rates \cite{Rienacker2002, Chakrabarti2004, Sabine2013, Klapp2013}. Indeed, little is known about active turbulence in 3D, even in the absence of shear.

By analysing the response of passive nematics to external driving forces, we have found that spatial inhomogeneities in the angular velocity of the director field, either due to elastic forces blocking the rotation or spatially differing vorticity fields, gives rise to walls. Thus tumbling nematics have an intrinsic tendency to generate more walls than aligning nematics.

Our finding that vorticity gradients help in the formation of walls is relevant to active turbulence. In active turbulence the walls and topological defects that are formed are the sources of vorticity, which then diffuses generating vorticity gradients. This helps in the formation of new walls which then sustain the active turbulence. To demonstrate this we analysed active turbulent patterns as a function of the alignment parameter $\lambda$. Our observations are quantified in Fig.~\ref{fig:lambdaeffects}. In panel (a) the normalised spatial correlations of the nematic director field $\left \langle \mathbf{Q}(0) : \mathbf{Q}(R) \right \rangle/\left \langle \mathbf{Q}(0) : \mathbf{Q}(0) \right \rangle$ are plotted as a function of the radial distance $R$ for a range of values of $\lambda$. As $\lambda$ decreases, there is a clear reduction in the characteristic correlation length of the the director field. This indicates that more walls are formed as the tumbling tendency of the director field increases. The increased number of walls gives rise to more topological defects, as apparent in Fig.~\ref{fig:ndeflambda}.

To conclude, we have shown that in passive nematics walls can be generated due to boundaries or imposed nonlinear flow fields. The effects are more pronounced 
when the nematic is tumbling. The walls can be easily destabilised by perturbing them, say with channels of non-uniform cross section. Then walls disintegrate into topological defects producing a director field analogous to that observed in active turbulence. Subjecting active nematics to external driving invokes a direct competition between active forces and externally imposed forces. When the imposed driving is large, active nematics behave like passive nematics. When the forces are comparable the competition leads to wall alignment in a simple shear flow and to active turbulence being confined to the centre of the channel in Poiseuille flow.

\section{Acknowledgements}

We thank Amin Doostmohammadi, Tyler Shendruk, Matthew L. Blow for helpful discussions and acknowledge funding from the ERC Advanced Grant MiCE.

\bibliography{refe}

\providecommand{\noopsort}[1]{}\providecommand{\singleletter}[1]{#1}%
\begin{thebibliography}{44}%
\makeatletter
\providecommand \@ifxundefined [1]{%
 \@ifx{#1\undefined}
}%
\providecommand \@ifnum [1]{%
 \ifnum #1\expandafter \@firstoftwo
 \else \expandafter \@secondoftwo
 \fi
}%
\providecommand \@ifx [1]{%
 \ifx #1\expandafter \@firstoftwo
 \else \expandafter \@secondoftwo
 \fi
}%
\providecommand \natexlab [1]{#1}%
\providecommand \enquote  [1]{``#1''}%
\providecommand \bibnamefont  [1]{#1}%
\providecommand \bibfnamefont [1]{#1}%
\providecommand \citenamefont [1]{#1}%
\providecommand \href@noop [0]{\@secondoftwo}%
\providecommand \href [0]{\begingroup \@sanitize@url \@href}%
\providecommand \@href[1]{\@@startlink{#1}\@@href}%
\providecommand \@@href[1]{\endgroup#1\@@endlink}%
\providecommand \@sanitize@url [0]{\catcode `\\12\catcode `\$12\catcode
  `\&12\catcode `\#12\catcode `\^12\catcode `\_12\catcode `\%12\relax}%
\providecommand \@@startlink[1]{}%
\providecommand \@@endlink[0]{}%
\providecommand \url  [0]{\begingroup\@sanitize@url \@url }%
\providecommand \@url [1]{\endgroup\@href {#1}{\urlprefix }}%
\providecommand \urlprefix  [0]{URL }%
\providecommand \Eprint [0]{\href }%
\providecommand \doibase [0]{http://dx.doi.org/}%
\providecommand \selectlanguage [0]{\@gobble}%
\providecommand \bibinfo  [0]{\@secondoftwo}%
\providecommand \bibfield  [0]{\@secondoftwo}%
\providecommand \translation [1]{[#1]}%
\providecommand \BibitemOpen [0]{}%
\providecommand \bibitemStop [0]{}%
\providecommand \bibitemNoStop [0]{.\EOS\space}%
\providecommand \EOS [0]{\spacefactor3000\relax}%
\providecommand \BibitemShut  [1]{\csname bibitem#1\endcsname}%
\let\auto@bib@innerbib\@empty
\bibitem [{\citenamefont {Marchetti}\ \emph {et~al.}(2013)\citenamefont
  {Marchetti}, \citenamefont {Joanny}, \citenamefont {Ramaswamy}, \citenamefont
  {Liverpool}, \citenamefont {Prost}, \citenamefont {Rao},\ and\ \citenamefont
  {Simha}}]{Marchetti2013}%
  \BibitemOpen
  \bibfield  {author} {\bibinfo {author} {\bibfnamefont {M.~C.}\ \bibnamefont
  {Marchetti}}, \bibinfo {author} {\bibfnamefont {J.~F.}\ \bibnamefont
  {Joanny}}, \bibinfo {author} {\bibfnamefont {S.}~\bibnamefont {Ramaswamy}},
  \bibinfo {author} {\bibfnamefont {T.~B.}\ \bibnamefont {Liverpool}}, \bibinfo
  {author} {\bibfnamefont {J.}~\bibnamefont {Prost}}, \bibinfo {author}
  {\bibfnamefont {M.}~\bibnamefont {Rao}}, \ and\ \bibinfo {author}
  {\bibfnamefont {R.~A.}\ \bibnamefont {Simha}},\ }\href {\doibase
  10.1103/RevModPhys.85.1143} {\bibfield  {journal} {\bibinfo  {journal} {Rev.
  Mod. Phys.}\ }\textbf {\bibinfo {volume} {85}},\ \bibinfo {pages} {1143}
  (\bibinfo {year} {2013})}\BibitemShut {NoStop}%
\bibitem [{\citenamefont {Koch}\ and\ \citenamefont
  {Subramanian}(2011)}]{Ganesh2011}%
  \BibitemOpen
  \bibfield  {author} {\bibinfo {author} {\bibfnamefont {D.~L.}\ \bibnamefont
  {Koch}}\ and\ \bibinfo {author} {\bibfnamefont {G.}~\bibnamefont
  {Subramanian}},\ }\href {\doibase 10.1146/annurev-fluid-121108-145434}
  {\bibfield  {journal} {\bibinfo  {journal} {Annu. Rev. Fluid Mech.}\ }\textbf
  {\bibinfo {volume} {43}},\ \bibinfo {pages} {637} (\bibinfo {year}
  {2011})}\BibitemShut {NoStop}%
\bibitem [{\citenamefont {Ramaswamy}(2010)}]{Sriram2010}%
  \BibitemOpen
  \bibfield  {author} {\bibinfo {author} {\bibfnamefont {S.}~\bibnamefont
  {Ramaswamy}},\ }\href {\doibase 10.1146/annurev-conmatphys-070909-104101}
  {\bibfield  {journal} {\bibinfo  {journal} {Annu. Rev. Cond. Mat. Phys.}\
  }\textbf {\bibinfo {volume} {1}},\ \bibinfo {pages} {323} (\bibinfo {year}
  {2010})}\BibitemShut {NoStop}%
\bibitem [{\citenamefont {Schaller}\ \emph {et~al.}(2013)\citenamefont
  {Schaller}, \citenamefont {Schmoller}, \citenamefont {Karakose},
  \citenamefont {Hammerich}, \citenamefont {Maier},\ and\ \citenamefont
  {Bausch}}]{Bausch2013b}%
  \BibitemOpen
  \bibfield  {author} {\bibinfo {author} {\bibfnamefont {V.}~\bibnamefont
  {Schaller}}, \bibinfo {author} {\bibfnamefont {K.~M.}\ \bibnamefont
  {Schmoller}}, \bibinfo {author} {\bibfnamefont {E.}~\bibnamefont {Karakose}},
  \bibinfo {author} {\bibfnamefont {B.}~\bibnamefont {Hammerich}}, \bibinfo
  {author} {\bibfnamefont {M.}~\bibnamefont {Maier}}, \ and\ \bibinfo {author}
  {\bibfnamefont {A.~R.}\ \bibnamefont {Bausch}},\ }\href {\doibase
  10.1039/C3SM50506E} {\bibfield  {journal} {\bibinfo  {journal} {Soft Matter}\
  }\textbf {\bibinfo {volume} {9}},\ \bibinfo {pages} {7229} (\bibinfo {year}
  {2013})}\BibitemShut {NoStop}%
\bibitem [{\citenamefont {Sanchez}\ \emph {et~al.}(2012)\citenamefont
  {Sanchez}, \citenamefont {Chen}, \citenamefont {DeCamp}, \citenamefont
  {Heymann},\ and\ \citenamefont {Dogic}}]{Dogic2012}%
  \BibitemOpen
  \bibfield  {author} {\bibinfo {author} {\bibfnamefont {T.}~\bibnamefont
  {Sanchez}}, \bibinfo {author} {\bibfnamefont {D.~T.~N.}\ \bibnamefont
  {Chen}}, \bibinfo {author} {\bibfnamefont {S.~J.}\ \bibnamefont {DeCamp}},
  \bibinfo {author} {\bibfnamefont {M.}~\bibnamefont {Heymann}}, \ and\
  \bibinfo {author} {\bibfnamefont {Z.}~\bibnamefont {Dogic}},\ }\href
  {\doibase 10.1038/nature11591} {\bibfield  {journal} {\bibinfo  {journal}
  {Nature}\ }\textbf {\bibinfo {volume} {491}},\ \bibinfo {pages} {431}
  (\bibinfo {year} {2012})}\BibitemShut {NoStop}%
\bibitem [{\citenamefont {Wensink}\ \emph {et~al.}(2012)\citenamefont
  {Wensink}, \citenamefont {Dunkel}, \citenamefont {Heidenreich}, \citenamefont
  {Drescher}, \citenamefont {Goldstein}, \citenamefont {Lowen},\ and\
  \citenamefont {Yeomans}}]{Julia2012}%
  \BibitemOpen
  \bibfield  {author} {\bibinfo {author} {\bibfnamefont {H.~H.}\ \bibnamefont
  {Wensink}}, \bibinfo {author} {\bibfnamefont {J.}~\bibnamefont {Dunkel}},
  \bibinfo {author} {\bibfnamefont {S.}~\bibnamefont {Heidenreich}}, \bibinfo
  {author} {\bibfnamefont {K.}~\bibnamefont {Drescher}}, \bibinfo {author}
  {\bibfnamefont {R.~E.}\ \bibnamefont {Goldstein}}, \bibinfo {author}
  {\bibfnamefont {H.}~\bibnamefont {Lowen}}, \ and\ \bibinfo {author}
  {\bibfnamefont {J.~M.}\ \bibnamefont {Yeomans}},\ }\href {\doibase
  10.1073/pnas.1202032109} {\bibfield  {journal} {\bibinfo  {journal} {PNAS}\
  }\textbf {\bibinfo {volume} {109}},\ \bibinfo {pages} {14308} (\bibinfo
  {year} {2012})}\BibitemShut {NoStop}%
\bibitem [{\citenamefont {Sokolov}\ and\ \citenamefont
  {Aranson}(2012)}]{Aranson2012}%
  \BibitemOpen
  \bibfield  {author} {\bibinfo {author} {\bibfnamefont {A.}~\bibnamefont
  {Sokolov}}\ and\ \bibinfo {author} {\bibfnamefont {I.~S.}\ \bibnamefont
  {Aranson}},\ }\href {\doibase 10.1103/PhysRevLett.109.248109} {\bibfield
  {journal} {\bibinfo  {journal} {Phys. Rev. Lett.}\ }\textbf {\bibinfo
  {volume} {109}},\ \bibinfo {pages} {248109} (\bibinfo {year}
  {2012})}\BibitemShut {NoStop}%
\bibitem [{\citenamefont {Cavagna}\ and\ \citenamefont
  {Giardina}(2014)}]{Cavagna2014}%
  \BibitemOpen
  \bibfield  {author} {\bibinfo {author} {\bibfnamefont {A.}~\bibnamefont
  {Cavagna}}\ and\ \bibinfo {author} {\bibfnamefont {I.}~\bibnamefont
  {Giardina}},\ }\href {\doibase 10.1146/annurev-conmatphys-031113-133834}
  {\bibfield  {journal} {\bibinfo  {journal} {Annu. Rev. Cond. Mat. Phys.}\
  }\textbf {\bibinfo {volume} {5}},\ \bibinfo {pages} {183} (\bibinfo {year}
  {2014})},\ \Eprint
  {http://arxiv.org/abs/http://dx.doi.org/10.1146/annurev-conmatphys-031113-133834}
  {http://dx.doi.org/10.1146/annurev-conmatphys-031113-133834} \BibitemShut
  {NoStop}%
\bibitem [{\citenamefont {Masoud}\ and\ \citenamefont
  {Shelley}(2014)}]{Masoud2014}%
  \BibitemOpen
  \bibfield  {author} {\bibinfo {author} {\bibfnamefont {H.}~\bibnamefont
  {Masoud}}\ and\ \bibinfo {author} {\bibfnamefont {M.~J.}\ \bibnamefont
  {Shelley}},\ }\href {\doibase 10.1103/PhysRevLett.112.128304} {\bibfield
  {journal} {\bibinfo  {journal} {Phys. Rev. Lett.}\ }\textbf {\bibinfo
  {volume} {112}},\ \bibinfo {pages} {128304} (\bibinfo {year}
  {2014})}\BibitemShut {NoStop}%
\bibitem [{\citenamefont {Golestanian}(2012)}]{Ramin2012}%
  \BibitemOpen
  \bibfield  {author} {\bibinfo {author} {\bibfnamefont {R.}~\bibnamefont
  {Golestanian}},\ }\href {\doibase 10.1103/PhysRevLett.108.038303} {\bibfield
  {journal} {\bibinfo  {journal} {Phys. Rev. Lett.}\ }\textbf {\bibinfo
  {volume} {108}},\ \bibinfo {pages} {038303} (\bibinfo {year}
  {2012})}\BibitemShut {NoStop}%
\bibitem [{\citenamefont {Narayan}\ \emph {et~al.}(2007)\citenamefont
  {Narayan}, \citenamefont {Ramaswamy},\ and\ \citenamefont
  {Menon}}]{Narayan2007}%
  \BibitemOpen
  \bibfield  {author} {\bibinfo {author} {\bibfnamefont {V.}~\bibnamefont
  {Narayan}}, \bibinfo {author} {\bibfnamefont {S.}~\bibnamefont {Ramaswamy}},
  \ and\ \bibinfo {author} {\bibfnamefont {N.}~\bibnamefont {Menon}},\ }\href
  {\doibase 10.1126/science.1140414} {\bibfield  {journal} {\bibinfo  {journal}
  {Science}\ }\textbf {\bibinfo {volume} {317}},\ \bibinfo {pages} {105}
  (\bibinfo {year} {2007})}\BibitemShut {NoStop}%
\bibitem [{\citenamefont {Giomi}(2014)}]{Giomi2014arxiv}%
  \BibitemOpen
  \bibfield  {author} {\bibinfo {author} {\bibfnamefont {L.}~\bibnamefont
  {Giomi}},\ }\href {\doibase arXiv:1409.1555} {\bibfield  {journal} {\bibinfo
  {journal} {arXiv:1409.1555}\ } (\bibinfo {year} {2014}),\
  arXiv:1409.1555}\BibitemShut {NoStop}%
\bibitem [{\citenamefont {Thampi}\ \emph
  {et~al.}(2014{\natexlab{a}})\citenamefont {Thampi}, \citenamefont
  {Golestanian},\ and\ \citenamefont {Yeomans}}]{ourepl2014}%
  \BibitemOpen
  \bibfield  {author} {\bibinfo {author} {\bibfnamefont {S.~P.}\ \bibnamefont
  {Thampi}}, \bibinfo {author} {\bibfnamefont {R.}~\bibnamefont {Golestanian}},
  \ and\ \bibinfo {author} {\bibfnamefont {J.~M.}\ \bibnamefont {Yeomans}},\
  }\href {\doibase 10.1209/0295-5075/105/18001} {\bibfield  {journal} {\bibinfo
   {journal} {Europhys. Lett.}\ }\textbf {\bibinfo {volume} {105}},\ \bibinfo
  {pages} {18001} (\bibinfo {year} {2014}{\natexlab{a}})}\BibitemShut {NoStop}%
\bibitem [{\citenamefont {Giomi}\ \emph {et~al.}(2014)\citenamefont {Giomi},
  \citenamefont {Bowick}, \citenamefont {Mishra}, \citenamefont {Sknepnek},\
  and\ \citenamefont {Cristina~Marchetti}}]{Giomi2014}%
  \BibitemOpen
  \bibfield  {author} {\bibinfo {author} {\bibfnamefont {L.}~\bibnamefont
  {Giomi}}, \bibinfo {author} {\bibfnamefont {M.~J.}\ \bibnamefont {Bowick}},
  \bibinfo {author} {\bibfnamefont {P.}~\bibnamefont {Mishra}}, \bibinfo
  {author} {\bibfnamefont {R.}~\bibnamefont {Sknepnek}}, \ and\ \bibinfo
  {author} {\bibfnamefont {M.}~\bibnamefont {Cristina~Marchetti}},\ }\href
  {\doibase 10.1098/rsta.2013.0365} {\bibfield  {journal} {\bibinfo  {journal}
  {Philos. Trans. R. Soc. London, Ser. A}\ }\textbf {\bibinfo {volume} {372}}
  (\bibinfo {year} {2014}),\ 10.1098/rsta.2013.0365}\BibitemShut {NoStop}%
\bibitem [{\citenamefont {{Muhuri, S.}}\ \emph {et~al.}(2007)\citenamefont
  {{Muhuri, S.}}, \citenamefont {{Rao, M.}},\ and\ \citenamefont {{Ramaswamy,
  S.}}}]{Muhuri2007}%
  \BibitemOpen
  \bibfield  {author} {\bibinfo {author} {\bibnamefont {{Muhuri, S.}}},
  \bibinfo {author} {\bibnamefont {{Rao, M.}}}, \ and\ \bibinfo {author}
  {\bibnamefont {{Ramaswamy, S.}}},\ }\href {\doibase
  10.1209/0295-5075/78/48002} {\bibfield  {journal} {\bibinfo  {journal} {EPL}\
  }\textbf {\bibinfo {volume} {78}},\ \bibinfo {pages} {48002} (\bibinfo {year}
  {2007})}\BibitemShut {NoStop}%
\bibitem [{\citenamefont {de~Gennes}\ and\ \citenamefont
  {Prost}(1995)}]{DeGennesBook}%
  \BibitemOpen
  \bibfield  {author} {\bibinfo {author} {\bibfnamefont {P.~G.}\ \bibnamefont
  {de~Gennes}}\ and\ \bibinfo {author} {\bibfnamefont {J.}~\bibnamefont
  {Prost}},\ }\href@noop {} {\emph {\bibinfo {title} {The Physics of Liquid
  Crystals}}}\ (\bibinfo  {publisher} {Clarendon Press},\ \bibinfo {address}
  {Oxford},\ \bibinfo {year} {1995})\BibitemShut {NoStop}%
\bibitem [{\citenamefont {Beris}\ and\ \citenamefont
  {Edwards}(1994)}]{Berisbook}%
  \BibitemOpen
  \bibfield  {author} {\bibinfo {author} {\bibfnamefont {A.~N.}\ \bibnamefont
  {Beris}}\ and\ \bibinfo {author} {\bibfnamefont {B.~J.}\ \bibnamefont
  {Edwards}},\ }\href@noop {} {\emph {\bibinfo {title} {Thermodynamics of
  Flowing Systems}}}\ (\bibinfo  {publisher} {OUP},\ \bibinfo {address}
  {Oxford},\ \bibinfo {year} {1994})\BibitemShut {NoStop}%
\bibitem [{\citenamefont {Rien\"acker}\ \emph {et~al.}(2002)\citenamefont
  {Rien\"acker}, \citenamefont {Kr\"oger},\ and\ \citenamefont
  {Hess}}]{Rienacker2002}%
  \BibitemOpen
  \bibfield  {author} {\bibinfo {author} {\bibfnamefont {G.}~\bibnamefont
  {Rien\"acker}}, \bibinfo {author} {\bibfnamefont {M.}~\bibnamefont
  {Kr\"oger}}, \ and\ \bibinfo {author} {\bibfnamefont {S.}~\bibnamefont
  {Hess}},\ }\href {\doibase 10.1103/PhysRevE.66.040702} {\bibfield  {journal}
  {\bibinfo  {journal} {Phys. Rev. E}\ }\textbf {\bibinfo {volume} {66}},\
  \bibinfo {pages} {040702} (\bibinfo {year} {2002})}\BibitemShut {NoStop}%
\bibitem [{\citenamefont {A.~Simha}\ and\ \citenamefont
  {Ramaswamy}(2002)}]{Sriram2002}%
  \BibitemOpen
  \bibfield  {author} {\bibinfo {author} {\bibfnamefont {R.}~\bibnamefont
  {A.~Simha}}\ and\ \bibinfo {author} {\bibfnamefont {S.}~\bibnamefont
  {Ramaswamy}},\ }\href {\doibase 10.1103/PhysRevLett.89.058101} {\bibfield
  {journal} {\bibinfo  {journal} {Phys. Rev. Lett.}\ }\textbf {\bibinfo
  {volume} {89}},\ \bibinfo {pages} {058101} (\bibinfo {year}
  {2002})}\BibitemShut {NoStop}%
\bibitem [{\citenamefont {Denniston}\ \emph
  {et~al.}(2001{\natexlab{a}})\citenamefont {Denniston}, \citenamefont
  {Orlandini},\ and\ \citenamefont {Yeomans}}]{Denniston2001}%
  \BibitemOpen
  \bibfield  {author} {\bibinfo {author} {\bibfnamefont {C.}~\bibnamefont
  {Denniston}}, \bibinfo {author} {\bibfnamefont {E.}~\bibnamefont
  {Orlandini}}, \ and\ \bibinfo {author} {\bibfnamefont {J.~M.}\ \bibnamefont
  {Yeomans}},\ }\href {\doibase 10.1103/PhysRevE.63.056702} {\bibfield
  {journal} {\bibinfo  {journal} {Phys. Rev. E}\ }\textbf {\bibinfo {volume}
  {63}},\ \bibinfo {pages} {056702} (\bibinfo {year}
  {2001}{\natexlab{a}})}\BibitemShut {NoStop}%
\bibitem [{\citenamefont {Denniston}\ \emph {et~al.}(2004)\citenamefont
  {Denniston}, \citenamefont {Marenduzzo}, \citenamefont {Orlandini},\ and\
  \citenamefont {Yeomans}}]{Denniston2004}%
  \BibitemOpen
  \bibfield  {author} {\bibinfo {author} {\bibfnamefont {C.}~\bibnamefont
  {Denniston}}, \bibinfo {author} {\bibfnamefont {D.}~\bibnamefont
  {Marenduzzo}}, \bibinfo {author} {\bibfnamefont {E.}~\bibnamefont
  {Orlandini}}, \ and\ \bibinfo {author} {\bibfnamefont {J.~M.}\ \bibnamefont
  {Yeomans}},\ }\href {\doibase 10.1098/rsta.2004.1416} {\bibfield  {journal}
  {\bibinfo  {journal} {Phil. Trans. R. Soc. Lond. A}\ }\textbf {\bibinfo
  {volume} {362}},\ \bibinfo {pages} {1745} (\bibinfo {year}
  {2004})}\BibitemShut {NoStop}%
\bibitem [{\citenamefont {Marenduzzo}\ \emph {et~al.}(2007)\citenamefont
  {Marenduzzo}, \citenamefont {Orlandini}, \citenamefont {Cates},\ and\
  \citenamefont {Yeomans}}]{Davide2007}%
  \BibitemOpen
  \bibfield  {author} {\bibinfo {author} {\bibfnamefont {D.}~\bibnamefont
  {Marenduzzo}}, \bibinfo {author} {\bibfnamefont {E.}~\bibnamefont
  {Orlandini}}, \bibinfo {author} {\bibfnamefont {M.~E.}\ \bibnamefont
  {Cates}}, \ and\ \bibinfo {author} {\bibfnamefont {J.~M.}\ \bibnamefont
  {Yeomans}},\ }\href {\doibase 10.1103/PhysRevE.76.031921} {\bibfield
  {journal} {\bibinfo  {journal} {Phys. Rev. E}\ }\textbf {\bibinfo {volume}
  {76}},\ \bibinfo {pages} {031921} (\bibinfo {year} {2007})}\BibitemShut
  {NoStop}%
\bibitem [{\citenamefont {Thampi}\ \emph
  {et~al.}(2014{\natexlab{b}})\citenamefont {Thampi}, \citenamefont
  {Golestanian},\ and\ \citenamefont {Yeomans}}]{ourpta2014}%
  \BibitemOpen
  \bibfield  {author} {\bibinfo {author} {\bibfnamefont {S.~P.}\ \bibnamefont
  {Thampi}}, \bibinfo {author} {\bibfnamefont {R.}~\bibnamefont {Golestanian}},
  \ and\ \bibinfo {author} {\bibfnamefont {J.~M.}\ \bibnamefont {Yeomans}},\
  }\href {\doibase 10.1098/rsta.2013.0366} {\bibfield  {journal} {\bibinfo
  {journal} {Philos. Trans. R. Soc. London, Ser. A}\ }\textbf {\bibinfo
  {volume} {372}} (\bibinfo {year} {2014}{\natexlab{b}}),\
  10.1098/rsta.2013.0366}\BibitemShut {NoStop}%
\bibitem [{\citenamefont {Voituriez}\ \emph {et~al.}(2005)\citenamefont
  {Voituriez}, \citenamefont {Joanny},\ and\ \citenamefont
  {Prost}}]{Joanny2005}%
  \BibitemOpen
  \bibfield  {author} {\bibinfo {author} {\bibfnamefont {R.}~\bibnamefont
  {Voituriez}}, \bibinfo {author} {\bibfnamefont {J.~F.}\ \bibnamefont
  {Joanny}}, \ and\ \bibinfo {author} {\bibfnamefont {J.}~\bibnamefont
  {Prost}},\ }\href {\doibase 10.1209/epl/i2004-10501-2} {\bibfield  {journal}
  {\bibinfo  {journal} {Europhys. Lett.}\ }\textbf {\bibinfo {volume} {70}},\
  \bibinfo {pages} {404} (\bibinfo {year} {2005})}\BibitemShut {NoStop}%
\bibitem [{\citenamefont {Fielding}\ \emph {et~al.}(2011)\citenamefont
  {Fielding}, \citenamefont {Marenduzzo},\ and\ \citenamefont
  {Cates}}]{Suzanne2011}%
  \BibitemOpen
  \bibfield  {author} {\bibinfo {author} {\bibfnamefont {S.~M.}\ \bibnamefont
  {Fielding}}, \bibinfo {author} {\bibfnamefont {D.}~\bibnamefont
  {Marenduzzo}}, \ and\ \bibinfo {author} {\bibfnamefont {M.~E.}\ \bibnamefont
  {Cates}},\ }\href {\doibase 10.1103/PhysRevE.83.041910} {\bibfield  {journal}
  {\bibinfo  {journal} {Phys. Rev. E}\ }\textbf {\bibinfo {volume} {83}},\
  \bibinfo {pages} {041910} (\bibinfo {year} {2011})}\BibitemShut {NoStop}%
\bibitem [{\citenamefont {Wagner}\ and\ \citenamefont
  {Pagonabarraga}(2002)}]{Wagner2002}%
  \BibitemOpen
  \bibfield  {author} {\bibinfo {author} {\bibfnamefont {A.}~\bibnamefont
  {Wagner}}\ and\ \bibinfo {author} {\bibfnamefont {I.}~\bibnamefont
  {Pagonabarraga}},\ }\href {\doibase 10.1023/A:1014595628808} {\bibfield
  {journal} {\bibinfo  {journal} {J. of Stat. Phys.}\ }\textbf {\bibinfo
  {volume} {107}},\ \bibinfo {pages} {521} (\bibinfo {year}
  {2002})}\BibitemShut {NoStop}%
\bibitem [{\citenamefont {Succi}(2013)}]{SucciBook}%
  \BibitemOpen
  \bibfield  {author} {\bibinfo {author} {\bibfnamefont {S.}~\bibnamefont
  {Succi}},\ }\href@noop {} {\emph {\bibinfo {title} {The Lattice Boltzmann
  Equation: For Fluid Dynamics and Beyond}}},\ Numerical Mathematics and
  Scientific Computation\ (\bibinfo  {publisher} {OUP},\ \bibinfo {address}
  {Oxford},\ \bibinfo {year} {2013})\BibitemShut {NoStop}%
\bibitem [{\citenamefont {Guo}\ \emph {et~al.}(2002)\citenamefont {Guo},
  \citenamefont {Zheng},\ and\ \citenamefont {Shi}}]{Guo2002}%
  \BibitemOpen
  \bibfield  {author} {\bibinfo {author} {\bibfnamefont {Z.}~\bibnamefont
  {Guo}}, \bibinfo {author} {\bibfnamefont {C.}~\bibnamefont {Zheng}}, \ and\
  \bibinfo {author} {\bibfnamefont {B.}~\bibnamefont {Shi}},\ }\href {\doibase
  10.1103/PhysRevE.65.046308} {\bibfield  {journal} {\bibinfo  {journal} {Phys.
  Rev. E}\ }\textbf {\bibinfo {volume} {65}},\ \bibinfo {pages} {046308}
  (\bibinfo {year} {2002})}\BibitemShut {NoStop}%
\bibitem [{\citenamefont {Cates}\ \emph {et~al.}(2008)\citenamefont {Cates},
  \citenamefont {Fielding}, \citenamefont {Marenduzzo}, \citenamefont
  {Orlandini},\ and\ \citenamefont {Yeomans}}]{Cates2008}%
  \BibitemOpen
  \bibfield  {author} {\bibinfo {author} {\bibfnamefont {M.~E.}\ \bibnamefont
  {Cates}}, \bibinfo {author} {\bibfnamefont {S.~M.}\ \bibnamefont {Fielding}},
  \bibinfo {author} {\bibfnamefont {D.}~\bibnamefont {Marenduzzo}}, \bibinfo
  {author} {\bibfnamefont {E.}~\bibnamefont {Orlandini}}, \ and\ \bibinfo
  {author} {\bibfnamefont {J.~M.}\ \bibnamefont {Yeomans}},\ }\href {\doibase
  10.1103/PhysRevLett.101.068102} {\bibfield  {journal} {\bibinfo  {journal}
  {Phys. Rev. Lett.}\ }\textbf {\bibinfo {volume} {101}},\ \bibinfo {pages}
  {068102} (\bibinfo {year} {2008})}\BibitemShut {NoStop}%
\bibitem [{\citenamefont {Henrich}\ \emph {et~al.}(2010)\citenamefont
  {Henrich}, \citenamefont {Stratford}, \citenamefont {Marenduzzo},\ and\
  \citenamefont {Cates}}]{Henrich2010}%
  \BibitemOpen
  \bibfield  {author} {\bibinfo {author} {\bibfnamefont {O.}~\bibnamefont
  {Henrich}}, \bibinfo {author} {\bibfnamefont {K.}~\bibnamefont {Stratford}},
  \bibinfo {author} {\bibfnamefont {D.}~\bibnamefont {Marenduzzo}}, \ and\
  \bibinfo {author} {\bibfnamefont {M.~E.}\ \bibnamefont {Cates}},\ }\href
  {\doibase 10.1073/pnas.1004269107} {\bibfield  {journal} {\bibinfo  {journal}
  {PNAS}\ }\textbf {\bibinfo {volume} {107}},\ \bibinfo {pages} {13212}
  (\bibinfo {year} {2010})}\BibitemShut {NoStop}%
\bibitem [{\citenamefont {Thampi}\ \emph {et~al.}(2013)\citenamefont {Thampi},
  \citenamefont {Golestanian},\ and\ \citenamefont {Yeomans}}]{ourprl2013}%
  \BibitemOpen
  \bibfield  {author} {\bibinfo {author} {\bibfnamefont {S.~P.}\ \bibnamefont
  {Thampi}}, \bibinfo {author} {\bibfnamefont {R.}~\bibnamefont {Golestanian}},
  \ and\ \bibinfo {author} {\bibfnamefont {J.~M.}\ \bibnamefont {Yeomans}},\
  }\href {\doibase 10.1103/PhysRevLett.111.118101} {\bibfield  {journal}
  {\bibinfo  {journal} {Phys. Rev. Lett.}\ }\textbf {\bibinfo {volume} {111}},\
  \bibinfo {pages} {118101} (\bibinfo {year} {2013})}\BibitemShut {NoStop}%
\bibitem [{\citenamefont {Denniston}\ \emph
  {et~al.}(2001{\natexlab{b}})\citenamefont {Denniston}, \citenamefont
  {Orlandini},\ and\ \citenamefont {Yeomans}}]{Denniston2001b}%
  \BibitemOpen
  \bibfield  {author} {\bibinfo {author} {\bibfnamefont {C.}~\bibnamefont
  {Denniston}}, \bibinfo {author} {\bibfnamefont {E.}~\bibnamefont
  {Orlandini}}, \ and\ \bibinfo {author} {\bibfnamefont {J.~M.}\ \bibnamefont
  {Yeomans}},\ }\href {\doibase
  http://dx.doi.org/10.1016/S1089-3156(01)00004-6} {\bibfield  {journal}
  {\bibinfo  {journal} {Comput. Theor. Polym. Sci.}\ }\textbf {\bibinfo
  {volume} {11}},\ \bibinfo {pages} {389 } (\bibinfo {year}
  {2001}{\natexlab{b}})}\BibitemShut {NoStop}%
\bibitem [{\citenamefont {Pasechnik}\ \emph {et~al.}(2009)\citenamefont
  {Pasechnik}, \citenamefont {Chigrinov},\ and\ \citenamefont
  {Shmeliova}}]{PasechnikBook}%
  \BibitemOpen
  \bibfield  {author} {\bibinfo {author} {\bibfnamefont {S.~V.}\ \bibnamefont
  {Pasechnik}}, \bibinfo {author} {\bibfnamefont {V.~G.}\ \bibnamefont
  {Chigrinov}}, \ and\ \bibinfo {author} {\bibfnamefont {D.~V.}\ \bibnamefont
  {Shmeliova}},\ }\href@noop {} {\emph {\bibinfo {title} {Liquid crystals:
  viscous and elastic properties in theory and applications}}}\ (\bibinfo
  {publisher} {John Wiley \& Sons},\ \bibinfo {address} {Weinheim},\ \bibinfo
  {year} {2009})\BibitemShut {NoStop}%
\bibitem [{\citenamefont {Batista}\ \emph {et~al.}(2014)\citenamefont
  {Batista}, \citenamefont {Blow},\ and\ \citenamefont {Gama}}]{Matthewarxiv}%
  \BibitemOpen
  \bibfield  {author} {\bibinfo {author} {\bibfnamefont {V.~M.~O.}\
  \bibnamefont {Batista}}, \bibinfo {author} {\bibfnamefont {M.~L.}\
  \bibnamefont {Blow}}, \ and\ \bibinfo {author} {\bibfnamefont {M.~M.~T.}\
  \bibnamefont {Gama}},\ }\href {\doibase arXiv:1409.4625} {\bibfield
  {journal} {\bibinfo  {journal} {arXiv:1409.4625}\ } (\bibinfo {year}
  {2014}),\ arXiv:1409.4625}\BibitemShut {NoStop}%
\bibitem [{\citenamefont {Sengupta}\ \emph
  {et~al.}(2013{\natexlab{a}})\citenamefont {Sengupta}, \citenamefont {Tkalec},
  \citenamefont {Ravnik}, \citenamefont {Yeomans}, \citenamefont {Bahr},\ and\
  \citenamefont {Herminghaus}}]{Anupam2013b}%
  \BibitemOpen
  \bibfield  {author} {\bibinfo {author} {\bibfnamefont {A.}~\bibnamefont
  {Sengupta}}, \bibinfo {author} {\bibfnamefont {U.}~\bibnamefont {Tkalec}},
  \bibinfo {author} {\bibfnamefont {M.}~\bibnamefont {Ravnik}}, \bibinfo
  {author} {\bibfnamefont {J.~M.}\ \bibnamefont {Yeomans}}, \bibinfo {author}
  {\bibfnamefont {C.}~\bibnamefont {Bahr}}, \ and\ \bibinfo {author}
  {\bibfnamefont {S.}~\bibnamefont {Herminghaus}},\ }\href {\doibase
  10.1103/PhysRevLett.110.048303} {\bibfield  {journal} {\bibinfo  {journal}
  {Phys. Rev. Lett.}\ }\textbf {\bibinfo {volume} {110}},\ \bibinfo {pages}
  {048303} (\bibinfo {year} {2013}{\natexlab{a}})}\BibitemShut {NoStop}%
\bibitem [{\citenamefont {Holmes}\ \emph {et~al.}(2010)\citenamefont {Holmes},
  \citenamefont {Cornford},\ and\ \citenamefont {Sambles}}]{Sambles2010}%
  \BibitemOpen
  \bibfield  {author} {\bibinfo {author} {\bibfnamefont {C.~J.}\ \bibnamefont
  {Holmes}}, \bibinfo {author} {\bibfnamefont {S.~L.}\ \bibnamefont
  {Cornford}}, \ and\ \bibinfo {author} {\bibfnamefont {J.~R.}\ \bibnamefont
  {Sambles}},\ }\href {\doibase 10.1103/PhysRevLett.104.248301} {\bibfield
  {journal} {\bibinfo  {journal} {Phys. Rev. Lett.}\ }\textbf {\bibinfo
  {volume} {104}},\ \bibinfo {pages} {248301} (\bibinfo {year}
  {2010})}\BibitemShut {NoStop}%
\bibitem [{\citenamefont {Lettinga}\ \emph {et~al.}(2005)\citenamefont
  {Lettinga}, \citenamefont {Dogic}, \citenamefont {Wang},\ and\ \citenamefont
  {Vermant}}]{Lettinga2005}%
  \BibitemOpen
  \bibfield  {author} {\bibinfo {author} {\bibfnamefont {M.~P.}\ \bibnamefont
  {Lettinga}}, \bibinfo {author} {\bibfnamefont {Z.}~\bibnamefont {Dogic}},
  \bibinfo {author} {\bibfnamefont {H.}~\bibnamefont {Wang}}, \ and\ \bibinfo
  {author} {\bibfnamefont {J.}~\bibnamefont {Vermant}},\ }\href {\doibase
  10.1021/la050116e} {\bibfield  {journal} {\bibinfo  {journal} {Langmuir}\
  }\textbf {\bibinfo {volume} {21}},\ \bibinfo {pages} {8048} (\bibinfo {year}
  {2005})},\ \bibinfo {note} {pMID: 16089419}\BibitemShut {NoStop}%
\bibitem [{\citenamefont {Sengupta}\ \emph {et~al.}(2014)\citenamefont
  {Sengupta}, \citenamefont {Herminghaus},\ and\ \citenamefont
  {Bahr}}]{Anupam2014}%
  \BibitemOpen
  \bibfield  {author} {\bibinfo {author} {\bibfnamefont {A.}~\bibnamefont
  {Sengupta}}, \bibinfo {author} {\bibfnamefont {S.}~\bibnamefont
  {Herminghaus}}, \ and\ \bibinfo {author} {\bibfnamefont {C.}~\bibnamefont
  {Bahr}},\ }\href {\doibase 10.1080/21680396.2014.963716} {\bibfield
  {journal} {\bibinfo  {journal} {Liq. Cryst. Rev.}\ }\textbf {\bibinfo
  {volume} {2}},\ \bibinfo {pages} {73} (\bibinfo {year} {2014})}\BibitemShut
  {NoStop}%
\bibitem [{\citenamefont {Sengupta}\ \emph {et~al.}(2011)\citenamefont
  {Sengupta}, \citenamefont {Tkalec},\ and\ \citenamefont {Bahr}}]{Anupam2011}%
  \BibitemOpen
  \bibfield  {author} {\bibinfo {author} {\bibfnamefont {A.}~\bibnamefont
  {Sengupta}}, \bibinfo {author} {\bibfnamefont {U.}~\bibnamefont {Tkalec}}, \
  and\ \bibinfo {author} {\bibfnamefont {C.}~\bibnamefont {Bahr}},\ }\href
  {\doibase 10.1039/C1SM05052D} {\bibfield  {journal} {\bibinfo  {journal}
  {Soft Matter}\ }\textbf {\bibinfo {volume} {7}},\ \bibinfo {pages} {6542}
  (\bibinfo {year} {2011})}\BibitemShut {NoStop}%
\bibitem [{\citenamefont {Sengupta}\ \emph
  {et~al.}(2013{\natexlab{b}})\citenamefont {Sengupta}, \citenamefont {Pieper},
  \citenamefont {Enderlein}, \citenamefont {Bahr},\ and\ \citenamefont
  {Herminghaus}}]{Anupam2013}%
  \BibitemOpen
  \bibfield  {author} {\bibinfo {author} {\bibfnamefont {A.}~\bibnamefont
  {Sengupta}}, \bibinfo {author} {\bibfnamefont {C.}~\bibnamefont {Pieper}},
  \bibinfo {author} {\bibfnamefont {J.}~\bibnamefont {Enderlein}}, \bibinfo
  {author} {\bibfnamefont {C.}~\bibnamefont {Bahr}}, \ and\ \bibinfo {author}
  {\bibfnamefont {S.}~\bibnamefont {Herminghaus}},\ }\href {\doibase
  10.1039/C2SM27337C} {\bibfield  {journal} {\bibinfo  {journal} {Soft Matter}\
  }\textbf {\bibinfo {volume} {9}},\ \bibinfo {pages} {1937} (\bibinfo {year}
  {2013}{\natexlab{b}})}\BibitemShut {NoStop}%
\bibitem [{\citenamefont {Sokolov}\ \emph {et~al.}(2007)\citenamefont
  {Sokolov}, \citenamefont {Aranson}, \citenamefont {Kessler},\ and\
  \citenamefont {Goldstein}}]{Sokolov2007}%
  \BibitemOpen
  \bibfield  {author} {\bibinfo {author} {\bibfnamefont {A.}~\bibnamefont
  {Sokolov}}, \bibinfo {author} {\bibfnamefont {I.}~\bibnamefont {Aranson}},
  \bibinfo {author} {\bibfnamefont {J.}~\bibnamefont {Kessler}}, \ and\
  \bibinfo {author} {\bibfnamefont {R.}~\bibnamefont {Goldstein}},\ }\href
  {\doibase 10.1103/PhysRevLett.98.158102} {\bibfield  {journal} {\bibinfo
  {journal} {Phys. Rev. Lett.}\ }\textbf {\bibinfo {volume} {98}},\ \bibinfo
  {pages} {158102} (\bibinfo {year} {2007})}\BibitemShut {NoStop}%
\bibitem [{\citenamefont {Chakrabarti}\ \emph {et~al.}(2004)\citenamefont
  {Chakrabarti}, \citenamefont {Das}, \citenamefont {Dasgupta}, \citenamefont
  {Ramaswamy},\ and\ \citenamefont {Sood}}]{Chakrabarti2004}%
  \BibitemOpen
  \bibfield  {author} {\bibinfo {author} {\bibfnamefont {B.}~\bibnamefont
  {Chakrabarti}}, \bibinfo {author} {\bibfnamefont {M.}~\bibnamefont {Das}},
  \bibinfo {author} {\bibfnamefont {C.}~\bibnamefont {Dasgupta}}, \bibinfo
  {author} {\bibfnamefont {S.}~\bibnamefont {Ramaswamy}}, \ and\ \bibinfo
  {author} {\bibfnamefont {A.}~\bibnamefont {Sood}},\ }\href {\doibase
  10.1103/PhysRevLett.92.055501} {\bibfield  {journal} {\bibinfo  {journal}
  {Phys. Rev. Lett.}\ }\textbf {\bibinfo {volume} {92}},\ \bibinfo {pages}
  {055501} (\bibinfo {year} {2004})}\BibitemShut {NoStop}%
\bibitem [{\citenamefont {Strehober}\ \emph
  {et~al.}(2013{\natexlab{a}})\citenamefont {Strehober}, \citenamefont
  {Engel},\ and\ \citenamefont {Klapp}}]{Sabine2013}%
  \BibitemOpen
  \bibfield  {author} {\bibinfo {author} {\bibfnamefont {D.~A.}\ \bibnamefont
  {Strehober}}, \bibinfo {author} {\bibfnamefont {H.}~\bibnamefont {Engel}}, \
  and\ \bibinfo {author} {\bibfnamefont {S.~H.~L.}\ \bibnamefont {Klapp}},\
  }\href {\doibase 10.1103/PhysRevE.88.012505} {\bibfield  {journal} {\bibinfo
  {journal} {Phys. Rev. E}\ }\textbf {\bibinfo {volume} {88}},\ \bibinfo
  {pages} {012505} (\bibinfo {year} {2013}{\natexlab{a}})}\BibitemShut
  {NoStop}%
\bibitem [{\citenamefont {Strehober}\ \emph
  {et~al.}(2013{\natexlab{b}})\citenamefont {Strehober}, \citenamefont
  {Sch\"oll},\ and\ \citenamefont {Klapp}}]{Klapp2013}%
  \BibitemOpen
  \bibfield  {author} {\bibinfo {author} {\bibfnamefont {D.~A.}\ \bibnamefont
  {Strehober}}, \bibinfo {author} {\bibfnamefont {E.}~\bibnamefont {Sch\"oll}},
  \ and\ \bibinfo {author} {\bibfnamefont {S.~H.~L.}\ \bibnamefont {Klapp}},\
  }\href {\doibase 10.1103/PhysRevE.88.062509} {\bibfield  {journal} {\bibinfo
  {journal} {Phys. Rev. E}\ }\textbf {\bibinfo {volume} {88}},\ \bibinfo
  {pages} {062509} (\bibinfo {year} {2013}{\natexlab{b}})}\BibitemShut
  {NoStop}%
\end{thebibliography}%

\end{document}